# Implementation and first application of EMC3-EIRENE on DTT for assessing the heat load on the ICRH antenna

H.S. Wu (吴昊声)[1], Y. Feng (冯玉和)[2], F. Subba[1], S. Ceccuzzi[34], P. Innocente[5], M.M. Robaldo[1], A.A. Tuccillo[6], R. Zanino[1]

1. NEMO Group, Dipartimento Energia, Politecnico di Torino, Corso Duca degli Abruzzi 24, 10129 Torino, Italy.
2. Max-Planck-Institut für Plasmaphysik, 17491 Greifswald, Germany.
3. ENEA, Nuclear Dept., Via Enrico Fermi 45, Frascati, 00044, Italy.
4. DTT S.c.a r.l., Via Enrico Fermi 45, Frascati, 00044, Italy.
5. Consorzio RFX (CNR, ENEA, INFN, Università di Padova, Acciaierie Venete SpA), Corso Stati Uniti 4, 35127 Padova, Italy.
6. CREATE Consortium, Via Claudio 21, Napoli, 80125, Italy.

## Abstract

This paper reports on the implementation process of the three-dimensional (3D) edge plasma transport code EMC3-EIRENE on the Divertor Tokamak Test (DTT) facility and the results of its first application, with a focus on assessing the heat load on the Ion Cyclotron Resonance Heating (ICRH) antenna surfaces. Using axisymmetric geometries and the SOLPS-ITER simulation results as a reference, we first evaluate the performance of the EMC3-EIRENE in describing axisymmetric plasmas to ensure correct code modelling setup for later complex 3D applications. We then incorporate the ICRH antenna structure in our 3D simulations, assuming different toroidal symmetry for the antenna to determine whether the heat load assessment can be carried out with reduced computational effort, and to what extent a 3D assessment deviates from a 2D approximation. The predictions of the 3D antenna heat load distribution are obtained and the peak values on the top plate and the two side plates reach 0.9, 2.1 and 3.8 MW/m$^2$, respectively. Finally, 3D gas puffing is included to evaluate its impact on the antenna heat load. The corresponding 3D edge plasma behavior is presented under the coupled effects of gas puffing and the antenna geometrical structure.

## 1. Introduction

The Divertor Tokamak Test (DTT) facility is a fusion experiment under construction in Italy, whose main mission is to explore alternative solutions for the power exhaust problem [1]. The Ion Cyclotron Resonance Heating (ICRH) system will provide a portion of the 45 MW auxiliary heating power required by DTT to achieve a power flux across the separatrix comparable to that expected in future fusion reactors, corresponding to $\frac{P_{sep}}{R} \sim 15\ MWm^{-1}$ [2]. Such conditions entail that plasma-facing components (PFCs) – especially those protruding from the first wall like ICRH antennas – are called to operate in a harsher environment compared to existing medium-size tokamaks. The DTT ICRH system is currently under design [3], and the antenna components in contact with the plasma require dedicated analyses to support the design process, the selection of PFC technology, and the optimization of both geometry and thermal management. To this aim, the three-dimensional (3D) heat load distribution on the antenna, arising from plasma–wall interactions under operational scenario, needs to be accurately assessed. This motivates the 3D edge plasma modeling for DTT, which must take into account the 3D antenna geometrical structures.

DTT is a tokamak with major radius $R$ = 2.19 m, minor radius $a$ = 0.7 m, toroidal magnetic field $B_T$ = 5.85 T, plasma current $I_p$ = 5.5 MA, and pulse length of 100 s. In the baseline configuration, 4 equatorial ports out of 18 are allocated to ICRH antennas, which can be upgraded to a maximum number of 6 according to the performance achieved during the first experimental years. According to radiofrequency simulations, the coupling of ICRH waves in DTT with fixed antennas flush with the first wall, i.e. at $R$ = 2.95 m, is highly ineffective [4]. To attain acceptable performance, a cantilevered, radially movable antenna structure is being conceived to move the antenna closer to the plasma when ICRH power is required. Moreover, owing to mechanical constraints, the poloidal shape of the antenna front face does not perfectly match DTT magnetic flux surfaces [4].

Compared with previous studies [5][6], which relied on the field-line tracing method with prescribed upstream parallel heat fluxes to simulate the heat flux on antenna, the state-of-the-art 3D edge plasma transport code EMC3-EIRENE[7][8] which describe the plasma in a self-consistent way is employed in the present work to predict the 3D heat load distribution on the ICRF antenna. This represents the first application of EMC3-EIRENE on DTT. EMC3-EIRENE includes a 3D macroscopic fluid plasma solver EMC3 [7] and a neutral particle transport solver EIRENE [9]. The EMC3-EIRENE code has been widely applied to stellarators and tokamaks for investigating plasma behavior and divertor performance under 3D conditions, e.g. magnetic configurations [10][11][12], geometrical structure [13] [14] and gas puffing [15].

A workflow including pre-processing, mesh generation and post-processing has been developed and applied for DTT 3D edge plasma modelling. To establish confidence in the modelling setup before addressing complex 3D effects, EMC3-EIRENE simulation results are first compared with SOLPS-ITER results for toroidally axisymmetric conditions.

This paper is structured as follows: Section 2 briefly introduces the development of the workflow. Section 3 introduces the benchmark between EMC3-EIRENE and SOLPS-ITER. Section 4 presents predicted antenna heat load together with the scan of antenna toroidal symmetry. Section 5 discusses the effect of perpendicular transport coefficients on antenna heat load. Section 6 investigates the 3D gas puffing effect on antenna heat load. The summary and outlook are provided in section 7.

## 2. A Workflow

Similar to [17][18], a workflow was independently developed for DTT 3D edge plasma modelling which support arbitrary tokamak configurations, e.g. limiter, single null, multiple X-points. All simulation cases presented in this study were generated and analyzed using this workflow. The workflow consists of three sequential stages, including pre-processing, mesh generation and post-processing. The main tasks of each stage are summarized below. Detailed descriptions of the numerical methods, programming data structure, algorithms etc. are beyond the scope of the present work and will be presented in a separate publication.

- **Pre-processing**: Geometry information, e.g. the first wall and divertor, and magnetic equilibrium are used as inputs for the pre-processing stage. Based on the equilibrium, the magnetic topology is identified, together with the corresponding O-point and X-point(s) coordinates. Subsequently, the information required for mesh generation, including the magnetic topology and associated parameters, is generated. The optimization for geometry and equilibrium is also performed in this stage.
- **Mesh generation**: The features include supporting artificial magnetic topology, as well as both 2D and 3D structured grid generation. Based on the inputs from pre-processing stage, the 2D plasma grid is generated that can be either aligned with the magnetic flux surfaces or constructed independently of them, according to user-defined requirements. The 3D plasma grid is constructed through magnetic field-line tracing based on the 2D plasma grid. The 3D neutral grid is formed by a set of 2D grids distributed along the toroidal direction, with each 2D grid corresponding to a toroidal slice. Unlike the plasma grid, the 3D neutral grids are not required to align to magnetic line and can therefore be defined with a high degree of geometrical flexibility to accommodate first-wall and divertor geometries. The grid can be converted into various formats to accommodate different edge plasma codes, such as the 2D grid with SONNET format used in SOLPS-ITER simulation.
- **Post-processing**: The outputs, e.g. the plasma parameters and neutral parameters, of EMC3-EIRENE are processed in this stage for data analysis and visualization. 1D, 2D and 3D representations of simulation results are supported.

## 3. Benchmark with SOLPS-ITER

Given the inherent complexity of 3D plasma transport, the influence of 3D geometrical structure, and the first application of EMC3-EIRENE to DTT, the present study begins with simulation under toroidally axisymmetric conditions, without antenna structure. Although the plasma configuration is toroidally axisymmetric, EMC3-EIRENE employs a 3D computational grid generated by tracing magnetic field lines. Consequently, the 3D plasma grid is non-strictly axisymmetric, in contrast to conventional 2D edge plasma codes such as SOLPS-ITER [18][19]. Therefore, comparison with SOLPS-ITER provides not only a benchmark between the two codes [20] but also an assessment of the capability of EMC3-EIRENE to accurately reproduce axisymmetric plasma conditions using a non-axisymmetric 3D computational grid. To ensure a meaningful comparison between EMC3-EIRENE and SOLPS-ITER, the modelling setups were chosen to be as consistent as possible and are summarized as follows:

- The computational grids are shown in Figure 1. Benefiting from the mesh generation described in Section 2, a 2D plasma grid with high orthogonality [21] was generated for the SOLPS-ITER simulations. The EMC3-EIRENE 3D plasma grid ranges from -12° to 12° in the toroidal direction and is constructed by tracing magnetic field lines starting from the poloidal cross-section at $\phi$=0° which corresponds to the 2D plasma grid employed in the SOLPS-ITER simulations. Unlike the plasma grids, the 2D and 3D neutral grids are generated independently but represent the same physical domain, extending to the first wall. In EMC3-EIRENE, an additional core domain is included to account for neutral particles penetrating across the core plasma boundary. The penetrated neutral particles is ionized in the additional core domain and return as ions at the innermost core plasma boundary.
- Both plasma solver B2.5(SOLPS-ITER) and EMC3 require magnetic field information. At each grid point of the 2D and 3D plasma grids, the magnetic field is calculated using cubic Hermite interpolation from the equilibrium data. This approach minimizes discrepancies arising from numerical errors in the magnetic-field calculations between the two codes.
- The directions of plasma current in SOLPS and EMC3 simulations are the same, which are in the counterclockwise direction when viewed from top. The toroidal magnetic field direction is in the favorable direction, corresponding to the clockwise direction when viewed from top. The directions of plasma current $I_p$, toroidal magnetic field $B_t$, poloidal magnetic field $B_p$ and total magnetic field B are shown in Figure 1.
- The equilibrium is a standard single null divertor configuration and the input power at the core boundary is 8MW that equally distributed on electron and ion [22].
- Core fueling is considered that the D+ particle flux at the core boundary is 1.1 $\times 10^{22}$ D+/s both for SOLPS-ITER and EMC3-ERIENE resulting in the separatrix density equals to $5.5\times10^{19}\ m^{-3}$.
- At the divertor targets, the Bohm boundary applies that in EMC3 $v = c_s = \sqrt{\frac{T_i+T_e}{m_i}}$ while in SOLPS $v \geq c_s = \sqrt{\frac{T_i+T_e}{m_i}}$. The sheath heat transmission coefficients $\gamma_e$=4.8 and $\gamma_i$=2.5 are used in the two simulations. At plasma boundaries of the Scrape-Off Layer (SOL) and Private Flux Region (PFR), the decay length (=3cm) boundary conditions apply for density, momentum, electron and ion temperature equations.
- The atomic reactions used in EMC3 and SOLPS are based on the Kotov-2008 model [23], with volumetric recombination reactions excluded to avoid numerical instability [24]. In addition, this study focuses on the antenna heat load, which is associate with the upstream conditions, rather than divertor target detachment. Moreover, the recent advancing implementation of molecule-assisted recombination in EMC3-EIRENE [25] differs from that in SOLPS-ITER. Benchmark activity including recombination is ongoing and will be presented in future.
- Same perpendicular transport coefficients are used in SOLPS-ITER and EMC3-EIRENE simulations that $D_{\perp} = 0.2\ m^2s^{-1}$ and $\chi_{i\perp} = \chi_{e\perp} = 0.6\ m^2s^{-1}$. It should be mentioned that these values are only for benchmarking purposes and are not the values foreseen in DTT experiments.
- The pumping is mimicked by the surface at the bottom of the divertor with pumping albedo 0.99.

- Neither drifts nor currents are included in SOLPS-ITER and EMC3-EIRENE, thus ion density equals electron density.

The comparison results are in Figure 2 and Figure 3. At the Outer Mid-plane (OMP), good agreement is obtained for both the electron density $n_e$ and electron temperature $T_e$ profiles as shown in Figure 2(a) and (b). The difference in $n_e$ is mainly concentrated in the region $0.01 \leq R - R_{sep} \leq 0.02m$ and remains within 20%, while the $T_e$ profiles are almost identical along the whole OMP. The Outer Target (OT) profiles of parallel particle flux $\Gamma_{\parallel}$ and parallel heat flux $q_{\parallel}$ are in Figure 2(c) and (d). Reasonable agreement is obtained between SOLPS-ITER and EMC3-EIRENE in terms of the profile shapes and peak values. The discrepancies in the peak values are approximately 15%. A small shift (0.02m) in the peak locations is also observed, with EMC3-EIRENE predicting the peaks slightly closer to the strike point. In the far-SOL region, where the antenna structure is considered in the subsequent simulations, the two codes also show good agreement. This further increases our confidence in the EMC3-EIRENE modelling results.

The 2D and 3D distributions of $n_e$ and $T_e$ are shown in Figure 3. For EMC3-EIRENE, the 2D distributions are at poloidal cross section ϕ=0°. Comparing the 2D distributions of $n_e$ and $T_e$, reasonable agreement between EMC3-EIRENE and SOLPS-ITER are obtained as in Figure 3 (a)(b)(d)(e). From the EMC3-EIRENE 3D distributions in Figure 3 (c)(f), it can be found there is perfect toroidal symmetry as our expectation. This indicates that EMC3-EIRENE can accurately reproduce axisymmetric plasma conditions using a non-axisymmetric 3D computational grid. The benchmark results demonstrate that the EMC3-EIRENE modeling setup is reliable and suitable for investigating 3D effects.

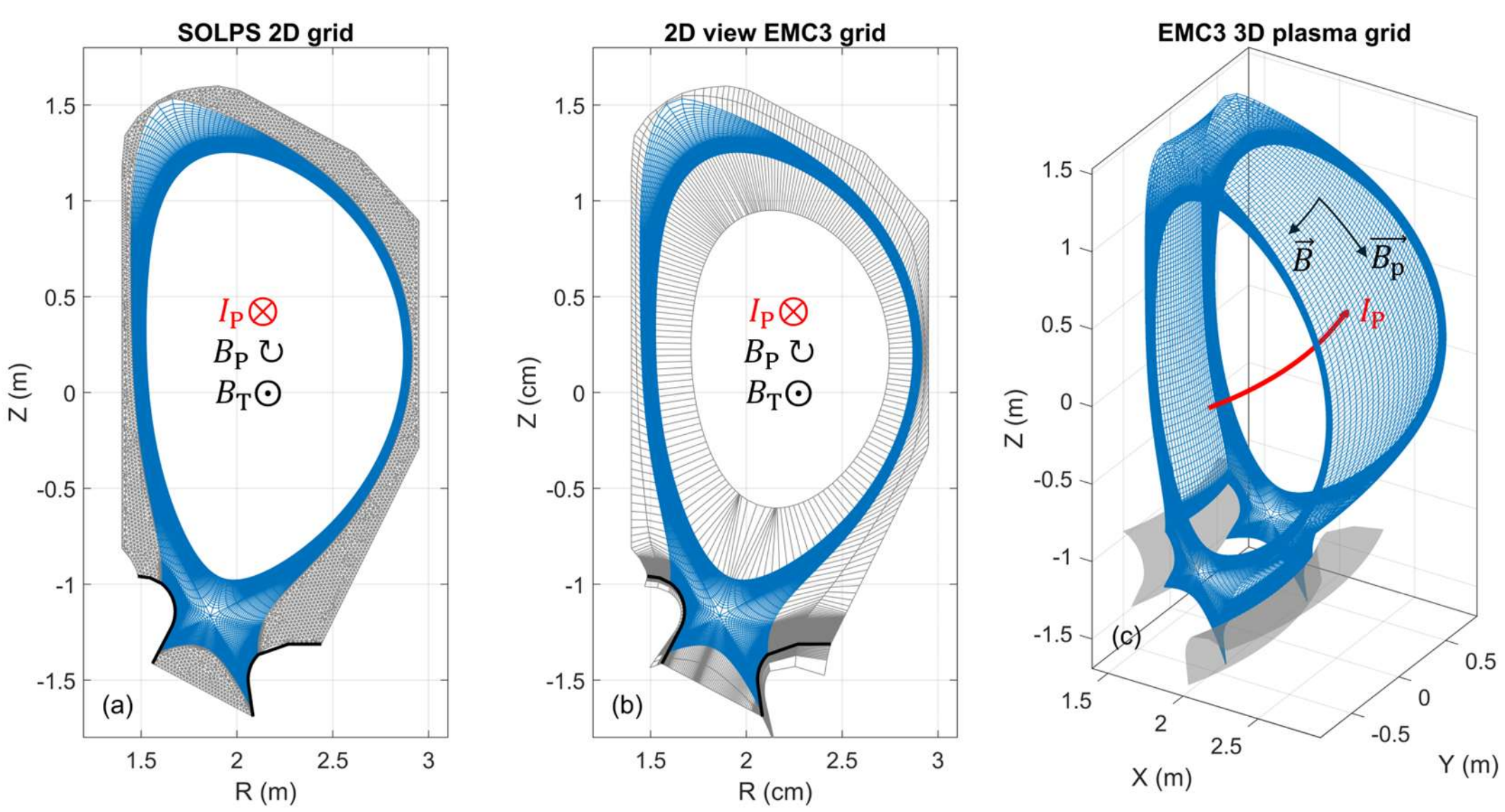

*Figure 1 (a) SOLPS-ITER 2D computational grid; (b) 2D poloidal cross-section view of EMC3-EIRENE 3D computational grid at φ=0° and (c) EMC3-EIRENE 3D plasma computational grid. In (a) and (b), the blue region denotes the plasma domain, while the gray region denotes the neutral domain. In (c) the divertor targets are plotted as gray surface. The directions of plasma current and magnetic field are marked.*

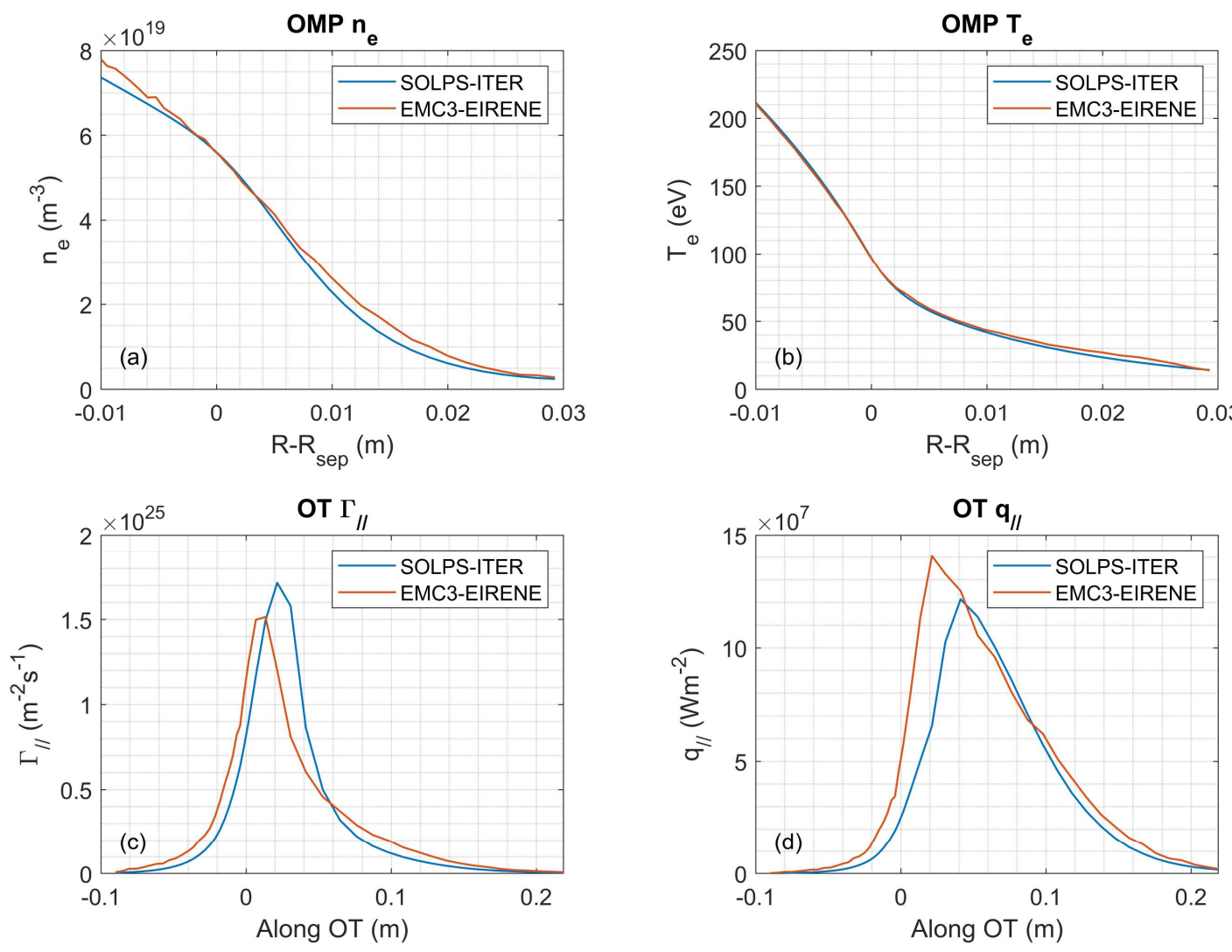

*Figure 2 Comparison of plasma profiles between SOLPS-ITER and EMC3-EIRENE. (a) Electron density $n_e$ profiles at the Outer Mid-plane; (b) Electron temperature $T_e$ profiles at the Outer Mid-plane, (c) parallel flux $\Gamma_{//}$ profiles at the Outer Target and (d) parallel heat flux $q_{//}$ profiles at the Outer Target.*

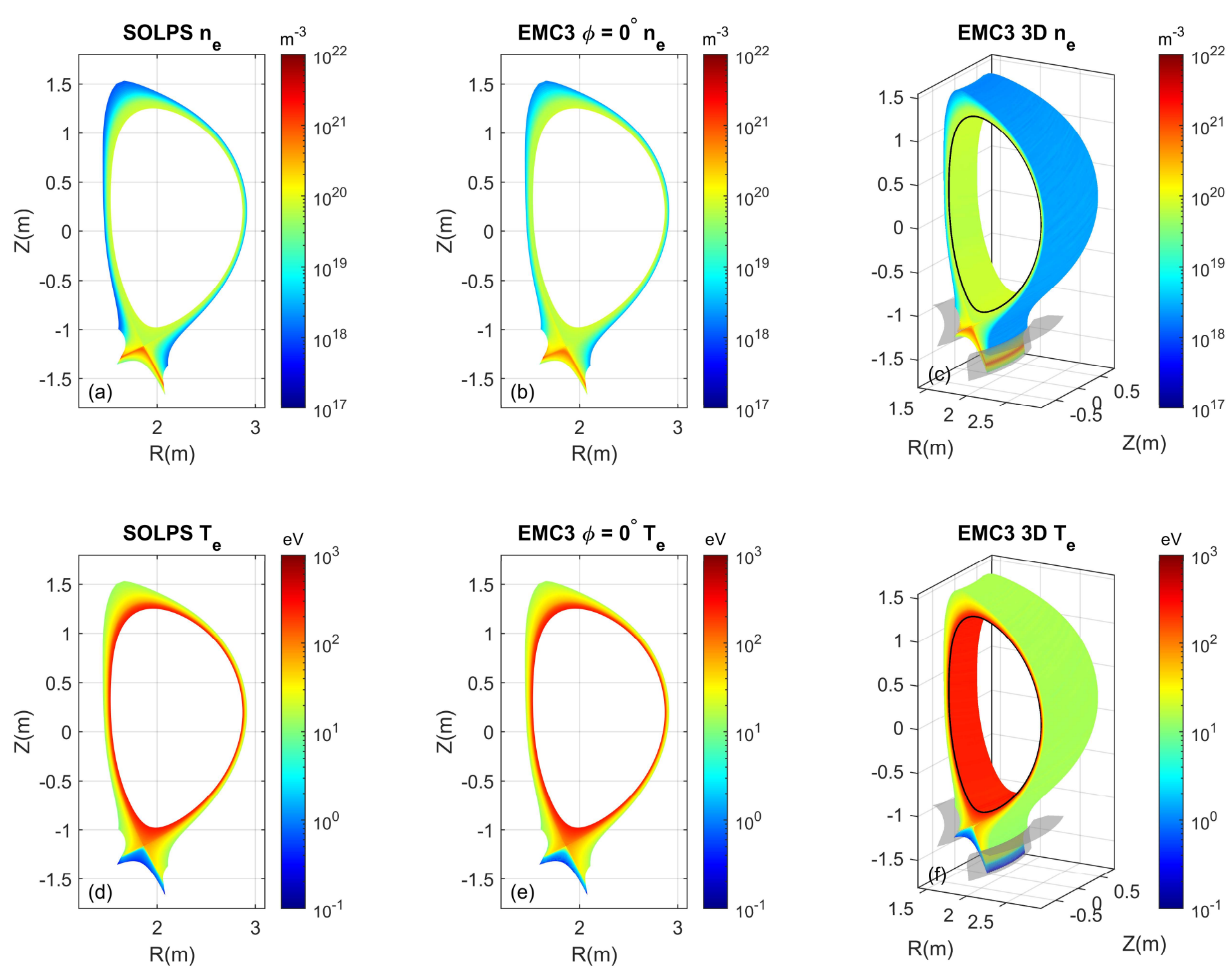

*Figure 3 Comparison of plasma distributions between SOLPS-ITER and EMC3-EIRENE. (a) SOLPS-ITER 2D electron density $n_e$ distribution; (b) EMC3-EIRENE 2D electron density $n_e$ distribution at $\phi$=0°; (c) EMC3-EIRENE 3D electron density $n_e$ distribution; (d) SOLPS-ITER 2D electron temperature $T_e$ distribution; (e) EMC3-EIRENE 2D electron temperature $T_e$ distribution at $\phi$=0° and (f) EMC3-EIRENE 3D electron temperature $T_e$ distribution.*

## 4. 3D Antenna heat load

The first ICRH antenna to be installed in DTT spans the toroidal range from 19.8° to 40.2° so that its total toroidal width is 20.4°. It could be equipped with three antenna structures that the final choice to be taken during the first years of operations [27]. In SOL, magnetic field lines intersect with the antenna structure. As a result, particles and energy could be transported along the field lines and deposited onto the antenna component surfaces resulting in plasma–wall interactions. Once the antenna structure is incorporated into the EMC3-EIRENE modelling, toroidal axisymmetric is no longer preserved, introducing 3D geometrical effect. In this section, we focus on predicting the heat load on the antenna surfaces and investigating how the 3D antenna geometry influences plasma behavior.

EMC3-EIRENE supports simulations with multiple toroidal zones within a specified toroidal range, allowing different toroidal symmetry (toroidal periodicity) conditions. This feature is utilized in this section to perform a toroidal symmetry scan which represents the number of antenna structures in the toroidal direction. From a physical perspective, this approach enables quantitative assessments of the 3D effect introduced by the antenna structure and providing a qualitative way to verify the simulation results. From a numerical perspective, it enables an assessment of the computational cost under different toroidal symmetry conditions, thereby providing guidance for the design and optimization of simulation strategies.

From a modeling perspective, the toroidal position is adjusted to the range from −10.2° to 10.2° with same toroidal width. The antenna geometry is modelled as a box with the plasma facing side (the front plate) shaped according to the real poloidal and toroidal profiles of the antenna front face, while all other sides are drawn as flat plates that the connection of plates is sharp without round-off corners. The latter choice allows for the prediction of the maximum expected parallel heat flux, which is pivotal information to define the antenna limiters and design their leading edges. The optimization of antenna structure is out of scope in this study. The 3D antenna geometrical structure is shown in Figure 4 (a), including top plate, front plate, side plate1 which is located at ϕ=-10.2° (blue) and side plate2 which located at ϕ=10.2° (red). Poloidal cross-section view at ϕ=0° are in Figure 4 (b) together with the computational grid. Grid cells that intersect the antenna structure are shown in gray, indicating that no plasma exists in these cells. These cells are mainly located in the far-SOL region. For example, at Z=0.2m, these cells are approximately 20 mm away from the separatrix at least, which is larger than the characteristic density decay length $\lambda_n$ (assuming $\lambda_n$ is 5-10mm). It should be noted that, in this study, the plasma domain is not extended to the first wall, meaning that the antenna structure is not fully encompassed by the plasma grid. The full domain simulation is ongoing and will be presented in future work.

We start from the simplest situation: a single toroidal zone ranging from -12° to 12° encompassing the antenna structure. Then, the number of toroidal zones, which do not include antenna structure, gradually increases to 3, 5, and 15. The number of antennas in each situation is represented by the antenna toroidal symmetry and are summarized as follows:

- $1 \times 24^{\circ}$zone with antenna toroidal symmetry of $\frac{360^{\circ}}{24^{\circ}} = 15$;
- $3 \times 24^{\circ}$domain with antenna toroidal symmetry of $\frac{360^{\circ}}{72^{\circ}} = 5$;
- $5 \times 24^{\circ}$domain with antenna toroidal symmetry of $\frac{360^{\circ}}{120^{\circ}} = 3$;
- $15 \times 24^{\circ}$domain covering the full $360^{\circ}$ toroidal domain, i.e., antenna toroidal symmetry equal to 1.

In this section, the DTT high power scenario [28] is considered: the input power at the core boundary is 30MW and the separatrix density $n_{e,sep} \geq 8 \times 10^{19} m^{-3}$. The perpendicular transport coefficients used in this section are $D_{\perp} = \frac{1}{3}\chi_{\perp} = 0.2\, m^2 s^{-1}$ that results in the reasonable characteristic decay lengths: $\lambda_n = 6.5\, mm, \lambda_T = 7.4\, mm, \lambda_q = 2.1\, mm$. The effect of perpendicular transport coefficients is addressed in section 4. It should be noted that impurity seeding is not considered in this study. This simplification is adopted to isolate the intrinsic effects of the 3D geometry on the heat load, without the additional complexity introduced by radiative cooling. Consequently, the predicted heat loads should be regarded as conservative upper-bound estimates, as impurity seeding would typically reduce the peak heat load through enhanced radiation in SOL. The inclusion of impurity seeding represents a meaningful extension of this work and will be explored in future studies.

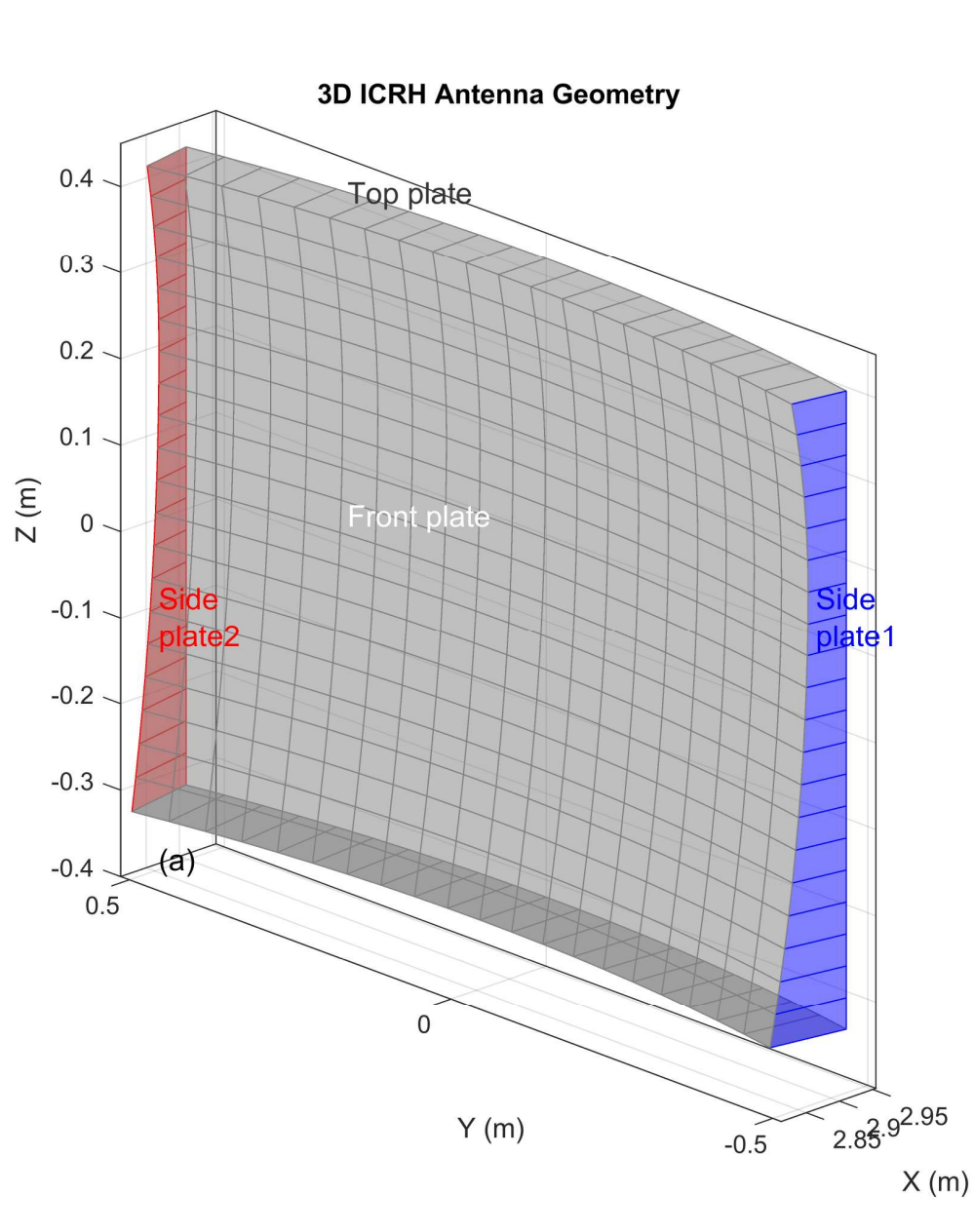

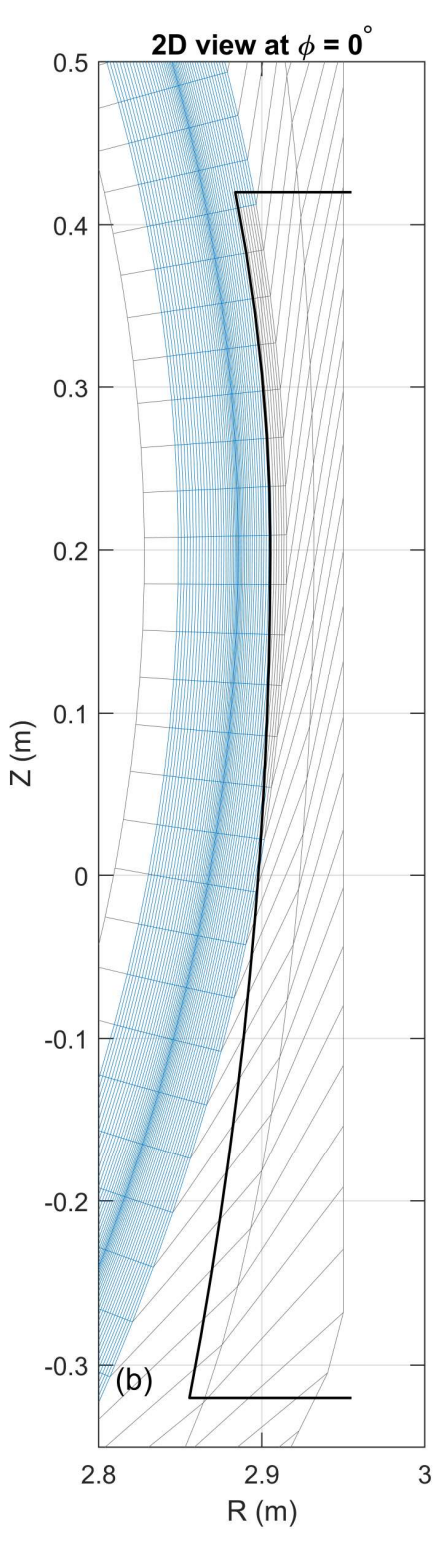

*Figure 4 (a) 3D view of the antenna geometrical structure, consisting of the top plate, front plate, side plate 1 (blue) at ϕ=-10.2° and side plate2 (red) at ϕ=10.2°. (b) Poloidal cross-section view of the antenna geometry at ϕ=0°. The antenna geometry is marked as black line, and the computational grid is also presented for reference. Plasma cells intersecting with the antenna are marked in gray that there is no plasma in such cells.*

The peak values of heat load on different antenna plates are summarized in Figure 5. The observed increase in peak heat loads with decreasing antenna toroid symmetry is physically reasonable. Larger toroidal symmetry provides a larger plasma-facing area for power deposition. Consequently, the energy fluxes are distributed over a larger surface area, leading to a reduction in the local peak heat load. A noteworthy result is observed for the case with a toroidal symmetry of 15 (the antenna structures cover about 300° of the full 360° toroidal domain), where the peak heat load on the side plates reaches approximately 1.5 MWm$^{-2}$. Although the total power deposited on each side plate is about 150 W, compared with approximately 600 W deposited on the front plate, the deposited power is concentrated over a much smaller surface area. Consequently, the local heat load on the side plate remains relatively high. However, it is unclear why the ratio of power deposited on the side and front plates, $\frac{150+150}{600} = 0.5$, is more than double the ratio of their toroidal widths $\frac{60^{\circ}}{300^{\circ}} = 0.2$, which will be investigated in the future.

EMC3-EIRENE employs a Monte Carlo method, and the CPU time required for each iteration depends on the number of test particles. Figure 6 shows the CPU time per iteration, normalized to 10,000 test particles. It can be found that the CPU-time for case with toroidal symmetry=1 is almost three times longer than the one in toroidal symmetry=3. This is reasonable because the computational domain is three times larger. However, Figure 5 indicates that the difference in the peak heat-load values between the toroidal symmetry=1 and toroidal symmetry=3 cases is less than 10%. This observation suggests an efficient modelling strategy: During the initial stage of modelling, when physical models are frequently adjusted, a toroidal symmetry of 3 is preferable because it reduces the computational cost substantially by considering only one-third of the toroidal domain, while introducing only a minor loss of accuracy. Once the modelling setup has been successfully established, simulations with a toroidal symmetry of 1 can be performed for more detailed analysis. This strategy has been applied in the section 5 and 6.

The 3D antenna heat load distribution for the case with toroidal symmetry=1 is shown in Figure 7. The heat load on the front plate is less than 0.2 MWm$^{-2}$ which is almost invisible in a linear scale color bar. The head load on the top plate has non-uniform distribution with the peak value io 0.9 $MWm^{-2}$ is located close to the side plate1. The heat load distributions on side plate1 and side plate2 have different patterns and the peak values are 2.1 $MWm^{-2}$ and 3.8 $MWm^{-2}$, respectively. The 3D antenna heat load distribution is a crucial input for the antenna design and ICRH cooling system definition.

The 3D plasma density distribution for the case with antenna toroidal symmetry=1 is shown in Figure 8, which illustrates the interaction between the plasma and the 3D antenna structure components. In SOL region, parallel transport dominates over perpendicular transport. The existence of antenna structure acts as a limiter and leads to plasma mainly deposited on the edge of side plates and rapidly decaying along the radial direction. Consequently, regions near the plasma boundary at the antenna structure exhibit lower plasma density, forming a characteristic "belt"-like structure on the magnetic surface as shown in Figure 8 (a). This is also confirmed from toroidal view in Figure 8 (d) where the plasma at boundary of the side plates (R~2.914m) have low density ~$1.0\times10^{17}$ $m^{-3}$. The effect of the 3D antenna structure on plasma density is localized in the far-SOL region as in Figure 8 (b)(c)(d) which presents the poloidal and toroidal cross-sections view of the plasma density at $\phi=0°$ plane and Z=0.2m plane, respectively. In the core and near-SOL regions, there is no obvious difference in the plasma density distribution.

The plasma density distributions for the cases with antenna toroidal symmetry=5 and 3 are shown in Figure 9 to future examine the effect introduced by the 3D antenna structure. From the 3D view of Figure 9 (a) and (c), together with Figure 8(a), it can be observed that as the antenna toroidal symmetry increases, the "belt"-like structures on the magnetic surfaces occurs more frequently in space. This trend confirms that the 3D edge plasma modeling successfully captures the effect introduced by the 3D antenna structure: higher antenna toroidal symmetry corresponds to a larger number of antennas in the full toroidal domain, leading to more intersections between the antenna structure and magnetic field lines. As shown in Figure 9 (b) and (d) which are the poloidal cross-section views of the plasma density at $\phi=0°$, variations along magnetic surfaces are observed in the far-SOL region e.g. top far-SOL where R~2.0m Z~1.4m, while in the core and near-SOL region there is no such variation. This indicates that three-dimensional structures protruding from the first wall can significantly affect plasma behavior in the far-SOL region. The assumption of toroidal symmetry used in 2D edge plasma modelling, especially when the simulated domain is extended to the first wall where 3D geometrical structures usually exist [29], may not be held for far-SOL plasmas and needs to be validated against experimental results.

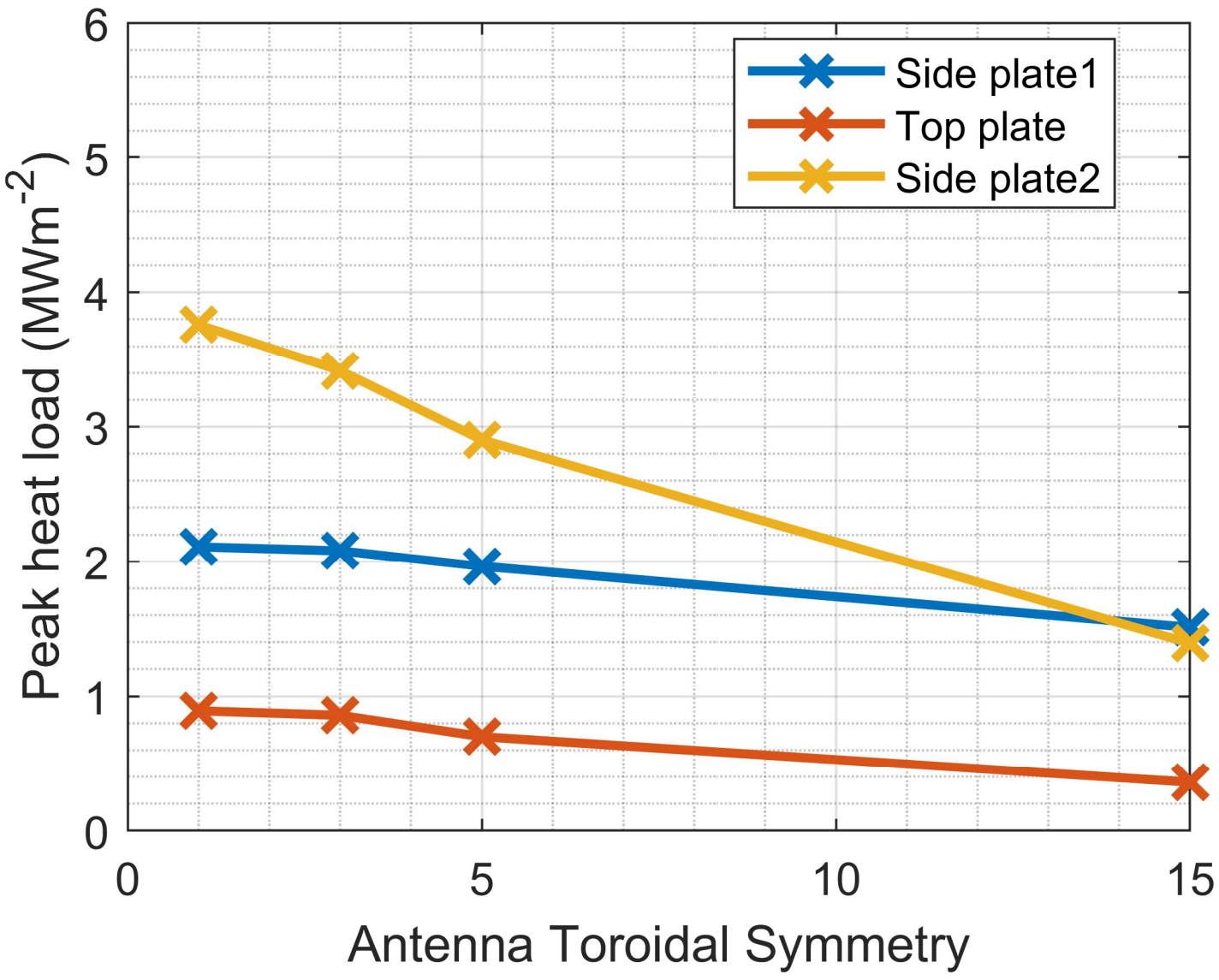

*Figure 5 The parametric scan of antenna toroidal symmetry: the peak values of antenna heat load at top and two side plates as a function of antenna toroidal symmetry. The antenna toroidal symmetry represents the number of antenna structures in the full toroidal domain.*

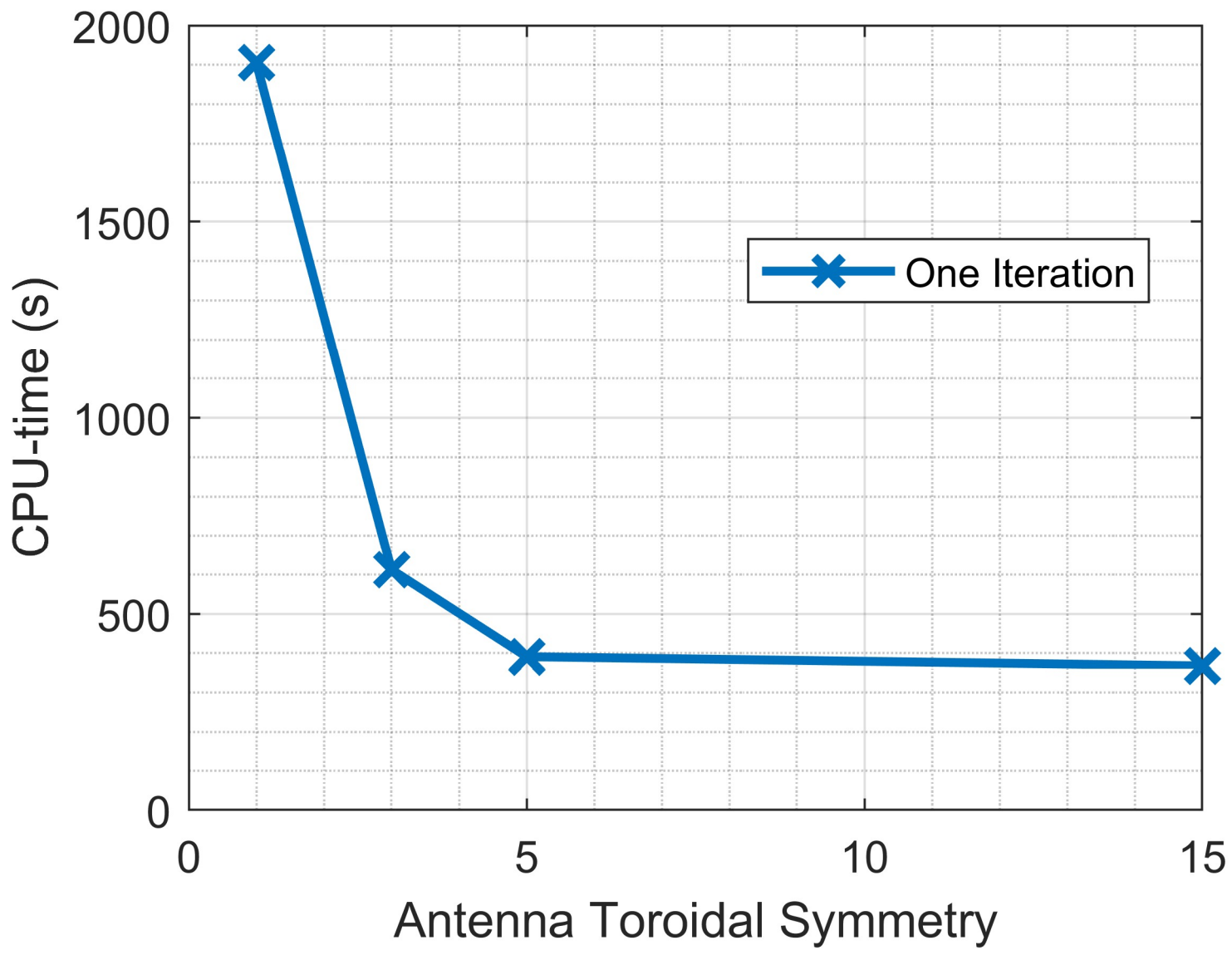

*Figure 6  CPU-time used for one iteration per 10,000 test particles.*

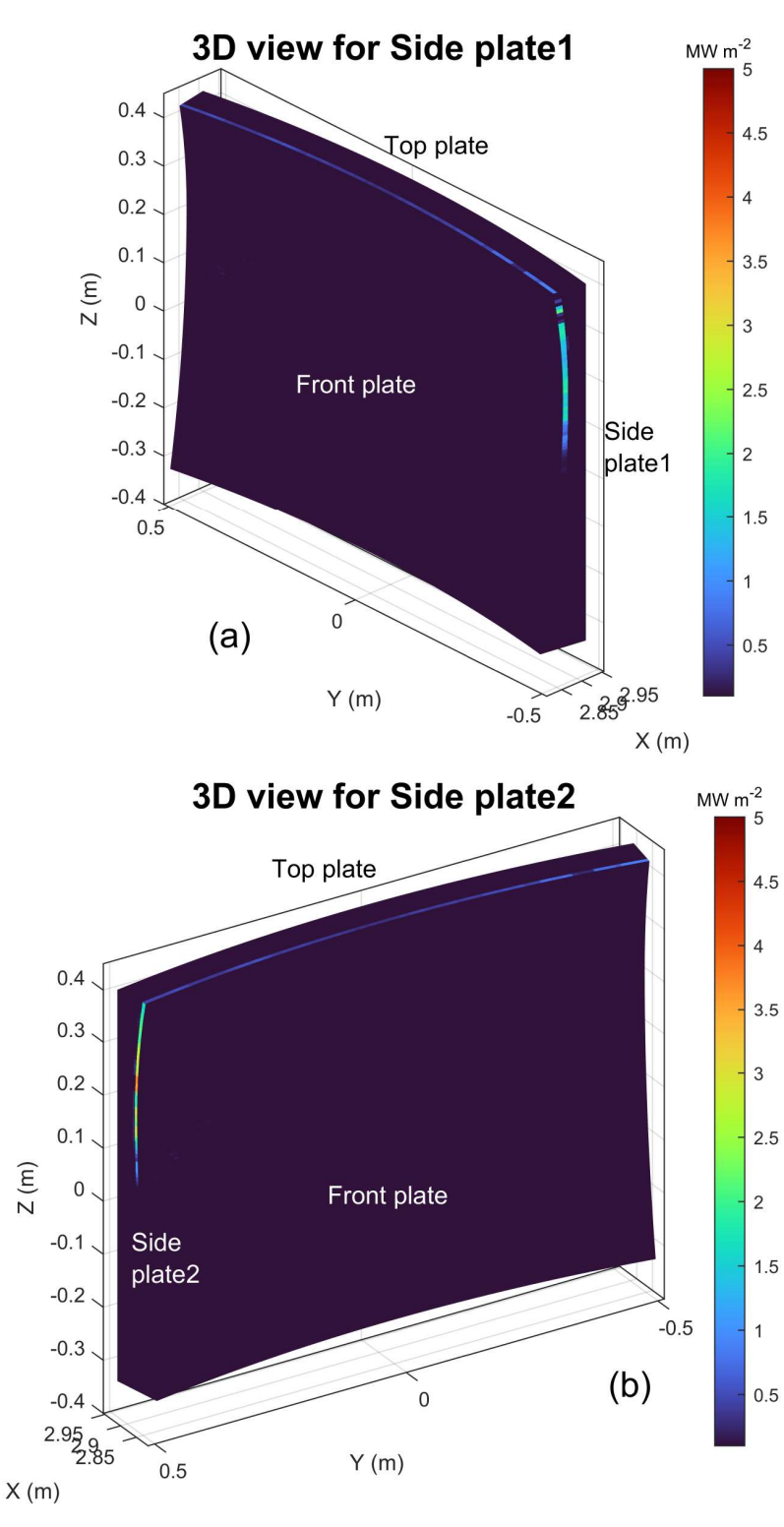

*Figure 7 ICRH Antenna 3D heat load distribution from different viewpoints: (a) view for side plate1 and (b) view for side plate2.*

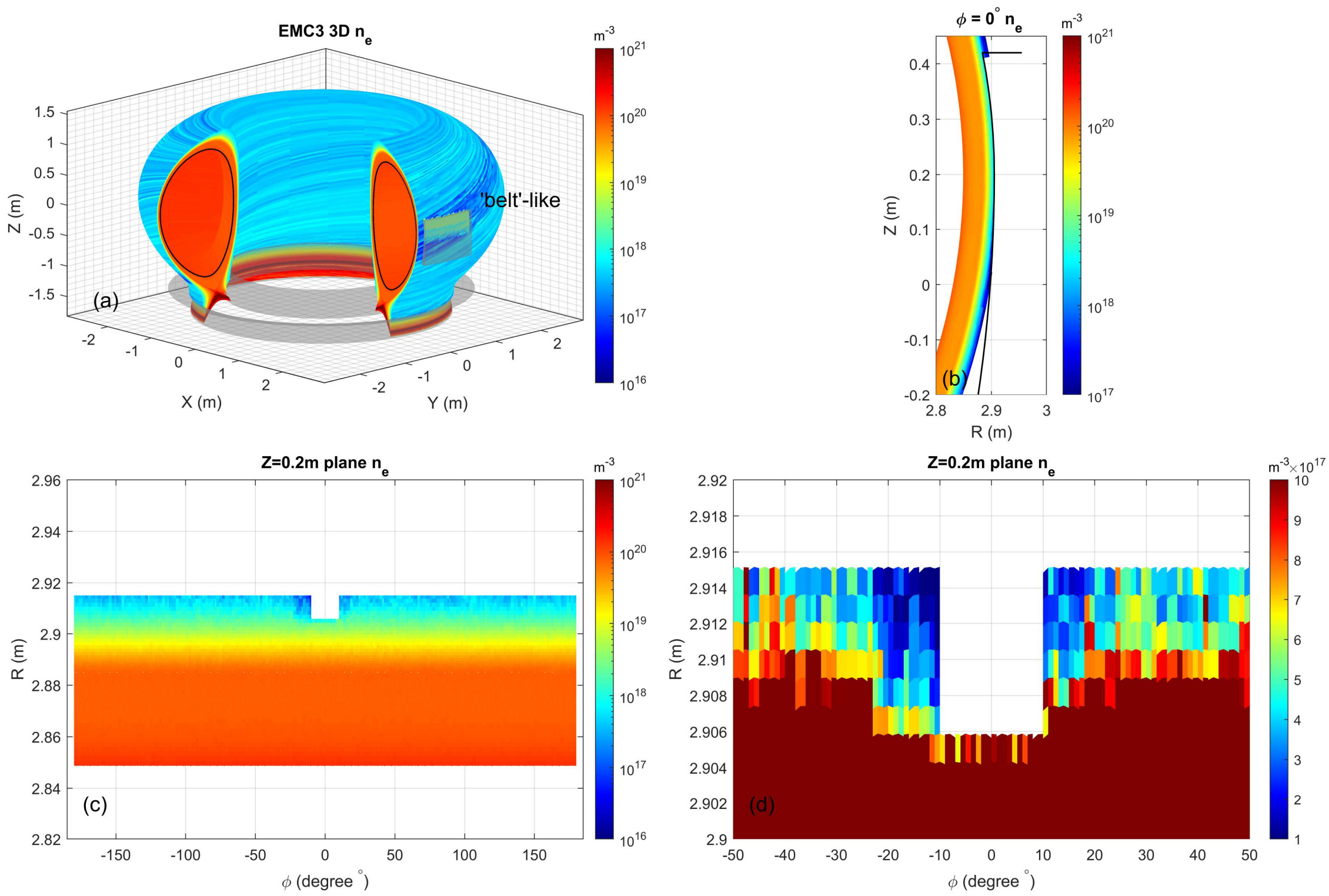

*Figure 8 Plasma density distributions for the antenna toroidal symmetry=1 case. (a) 3D view; (b) zoomed-in view near the antenna in the poloidal cross-section $\phi$=0°; (c) Toroidal cross-section view at Z=0.2m. (d) Zoomed-in view for the antenna structure in the toroidal cross-section Z=0.2 m. A linear scale is used to highlight the low-density region (~$1.0\times10^{17}$ $m^{-3}$) around the antenna side plates.*

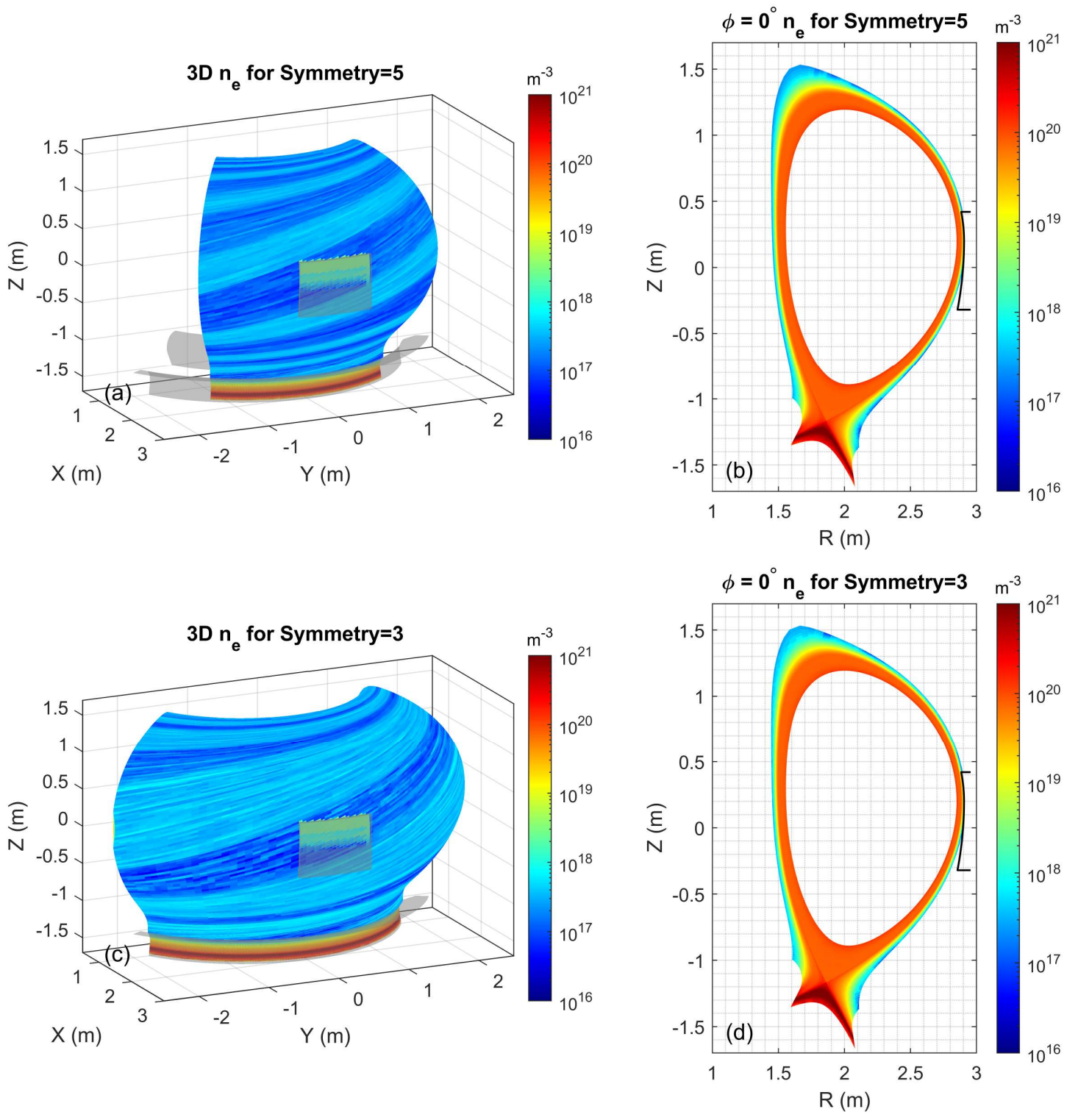

*Figure 9 Plasma density distributions. (a) 3D view and (b) 2D view at $\phi$=0° for the case of antenna toroidal symmetry=5; (c) 3D view and (d) 2D view at $\phi$=0° for the case of antenna toroidal symmetry=3;*

# 5. The Cross-field Transport Effect

In this study, the perpendicular transport coefficients are assumed to be constant values to represent cross-field transport. The cross-field transport coefficients affect the plasma parameters at the antenna plates, thereby governing particle and energy deposition. In this section, based on the case with antenna toroidal symmetry=1 in section 3, the effect of perpendicular transport coefficients is numerically investigated. The diffusion coefficient $D_{\perp}$is varied from 0.1 to 0.5 m$^{-2}$s$^{-1}$ with the assumption that $\chi_{\perp i} = \chi_{\perp e} = 3D_{\perp}$. The characteristic decay lengths $\lambda_n$, $\lambda_T$ for all cases are listed in Table 1. It can be found that as the $D_{\perp}$ and $\chi_{\perp}$ increase, the decay lengths of density and temperature increase by factors of 8 and 2, respectively. This suggests that cross-field transport has a stronger influence on the density profile than on the temperature profile, making $\lambda_n$ more sensitive to changes in the transport coefficient than $\lambda_T$.

*Table 1 Characteristic decay lengths for cases that transport coefficients $D_{\perp} = \frac{1}{3}\chi_{\perp}$ range from 0.1 to 0.5 $m^{-2}s^{-1}$. The electron density decay lengths $\lambda_n$ and electron temperature decay length $\lambda_T$ are included, which are measured at φ=-180° to avoid the effect of antenna structure.*

| Transport coefficients ($m^2s^{-1}$) | $\lambda_n$(mm) | $\lambda_T$(mm) |
|---|---|---|
| $D_{\perp} = \frac{1}{3}\chi_{\perp} = 0.1$ | 6.5 | 7.4 |
| $D_{\perp} = \frac{1}{3}\chi_{\perp} = 0.2$ | 22.6 | 9.7 |
| $D_{\perp} = \frac{1}{3}\chi_{\perp} = 0.3$ | 26.4 | 10.7 |
| $D_{\perp} = \frac{1}{3}\chi_{\perp} = 0.4$ | 42.1 | 13.4 |
| $D_{\perp} = \frac{1}{3}\chi_{\perp} = 0.5$ | 45.9 | 14.5 |

In this study, the plane at Z=0.2m is taken as the upstream location where antenna structure mainly intersects with plasma domain as in Figure 4(b). The upstream profiles at poloidal cross section ϕ=0° are presented in Figure 10. At comparable separatrix $n_e$, the density at the antenna front plate rises from 1.5×10$^{18}$ to 2.6×10$^{19}$ m$^{-3}$, increasing by approximately one order of magnitude with increasing $D_{\perp}$. There are noticeable differences in SOL density between the cases with $D_{\perp}$= 0.1 and 0.2-0.5. This can be explained by the ionization source due to the recycling neutral particles generated from the antenna front plate as shown in Figure 10 (c). It can be observed that the ionization source increases by at least one order of magnitude as $D_{\perp}$ increases. For the $T_e$ near the front plate, the values increase from 20 to 40 eV as the $\chi_{\perp}$ increase but remain at the same level.

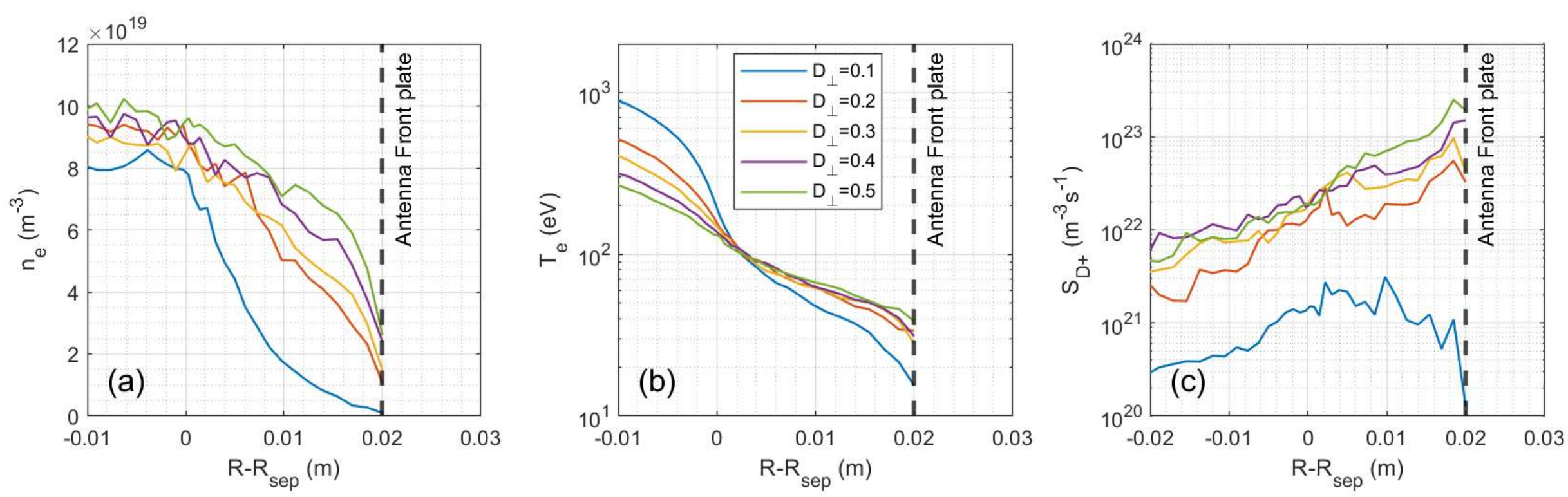

*Figure 10 Upstream (Z=0.02m) profiles of (a) electron density $n_e$ and (b) electron temperature $T_e$ and (c) D+ ionization source for the cases with the transport coefficients $D_{\perp} = \frac{1}{3}\chi_{\perp}$ ranging from 0.1 to 0.5 $m^2s^{-1}$ at poloidal cross section φ=0°. The values are cut off by the antenna front plate at R-$R_{sep}$=0.02m as shown in Figure 4.*

The heat load $q_{surf}$ on the antenna plates, where Bohm boundary condition is applied, includes the plasma contribution and surface recombination contribution. It can be written as the following equation:

$$q_{surf} = q_{\parallel} sin\theta = (q_{\parallel i} + q_{\parallel e} + \Gamma_{\parallel} E_{pot}) sin\theta = (\gamma_i n_i c_s T_i + \gamma_e n_e c_s T_e + n_i c_s \epsilon) sin\theta \quad (1)$$

where $\gamma_i = 2.5$ and $\gamma_e = 4.8$ are sheath heat transmission coefficients, $c_s = \sqrt{\frac{T_i+T_e}{m_i}}$ is plasma sound speed, $\epsilon = 15.6eV$ is the energy released to material component per ion due to recombination and $\theta$ is the tilting angle between the magnetic field and plate. The peak values of heat load $q_{surf}$ and the corresponding parallel heat flux $q_{\parallel}$ on different plates are presented in Figure 11. From Figure 11(a), it can be found that as the $D_{\perp}$ increase from 0.1 to 0.2, the peak values of heat load increase one order of magnitude higher e.g. from 2-4 MWm$^{-2}$ to 10-40 MWm$^{-2}$. Figure 10(b) indicates the Te remains at the same level that we could assume $T_i = T_e = constant$ in Equation (1). Thus $q_{surf} \propto n_i[\sqrt{\frac{2}{m_i}}(\gamma_i + \gamma_e)T^{\frac{3}{2}} + \epsilon T^{\frac{1}{2}}] \propto n_i$ indicates that the increase of heat load is attributed to the increase in density, which is consistent with Figure 10(a). Figure 11 (b) shows the peak values of $q_{\parallel}$ on different plates are closer than the those of $q_{surf}$. This indicates that the differences in heat load between the top plate and side plates are primarily caused by the local tilting angle instead of local plasma parameters. There is a notable difference between $q_{\parallel}$ on plate1 and that on plate2, with the peak value on plate2 being nearly double that on plate1. The asymmetric heat load indicates that, even under a toroidally axisymmetric magnetic equilibrium, different heat loads can still develop on the two side plates. This asymmetry is determined by the geometry of three-dimensional magnetic field and the antenna structure. We hypothesize that reversing plasma current would lead to the opposite pattern and will be verified in future study.

The 3D distributions of heat load are in Figure 12. The heat load on the front plates increases significantly from ~0.2-0.3 to 2-3 MWm$^{-2}$ as the increase of transport coefficients, which is consistent with the peak values in Figure 11. To the best of our knowledge, this work presents the first numerical prediction of 3D heat loads on the DTT antenna structures using EMC3-EIRENE. The predicted heat load distribution provides an important basis for the cooling system design of the antenna components.

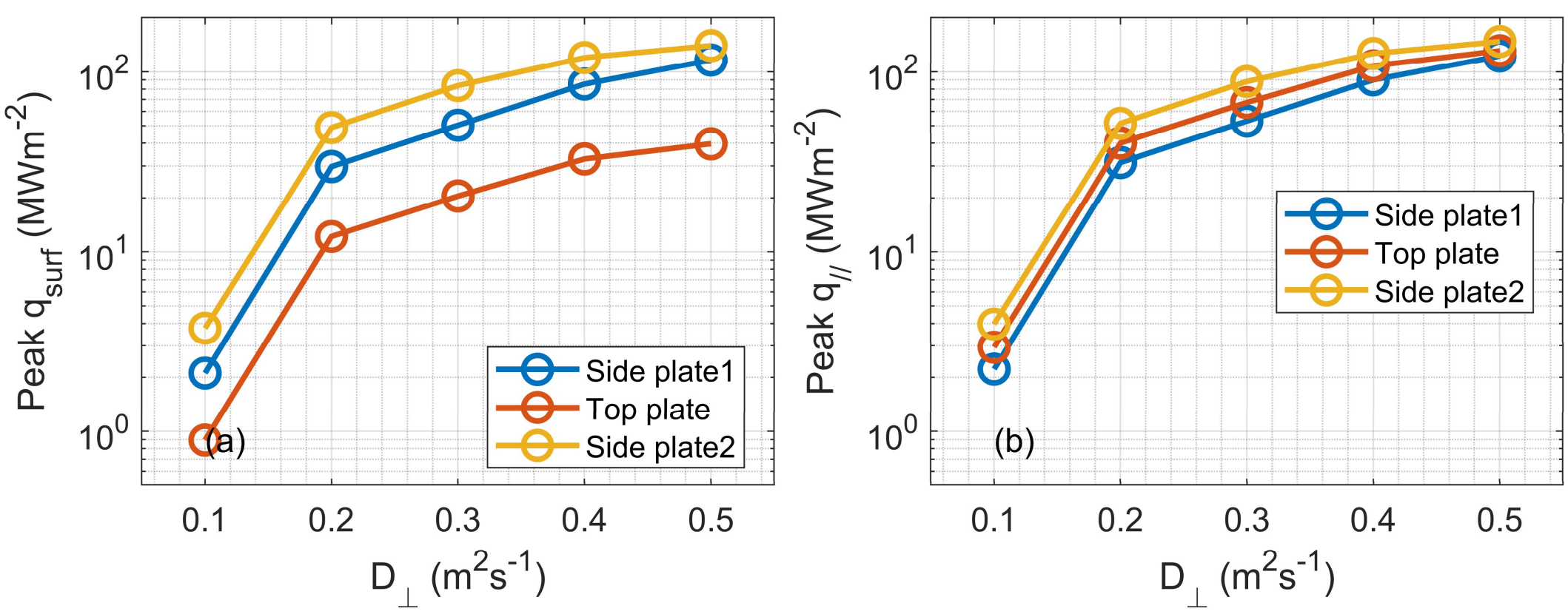

Figure 11 (a) Peak values of antenna heat load and (b) the corresponding parallel heat flux on side plate1, side plate2 and Top plate for the cases that transport coefficients $D_{\perp} = \frac{1}{3}\chi_{\perp}$ ranging from 0.1 to 0.5 $m^2s^{-1}$. A log scale is used for y-axis.

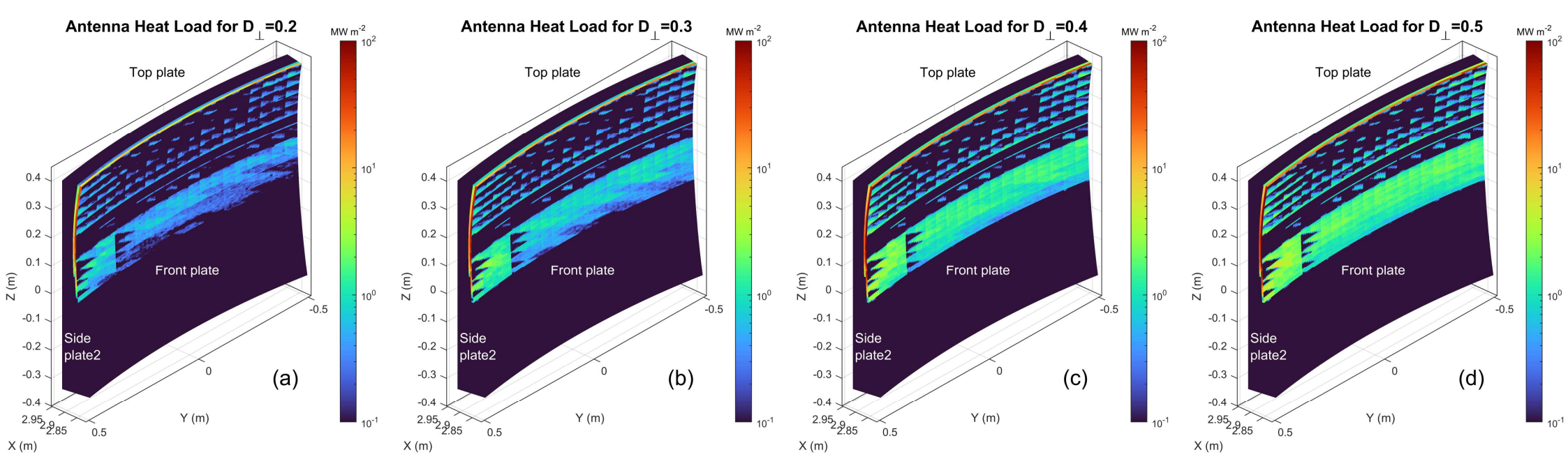

*Figure 12 3D heat load distributions of the ICRH antenna for cases with transport coefficients $D_{\perp} = \frac{1}{3}\chi_{\perp}$ ranging from 0.2 to 0.5 $m^2s^{-1}$, from (a) to (d). A log scale is used for the color bar to make the front plate patterns clear.*

## 6. 3D Gas Puffing Effect

In this section, we focus on numerically studying the impact of 3D gas puffing on the antenna heat load. The case which antenna toroidal symmetry=1 in Section 3, is treated as the reference case (without gas puffing). Based on the reference case, another four cases, named Case A, B, C, and D, with different deuterium gas puffing positions, are considered. The gas puffing is implemented by 3D point source. The positions have fixed R (=2.95 m) position and rate (=2.0×$10^{22}$ D/s) but different Z and toroidal positions are listed in Table 2. Compared to the reference case, the gas puffing positions are shifted toroidally in Case A and Case B. Relative to Case A, the gas puffing positions are shifted vertically in Case C and Case D.

*Table 2 The four cases with different gas puffing positions. The R (=2.95m) and rates (=2.0×$10^{22}$ D/s) are the same. The different Z and Φ positions are listed.*

| Case | Z (m) | Φ (º) |
|---|---|---|
| A | 0.00 | 20 |
| B | 0.00 | 40 |
| C | 0.45 | 20 |
| D | 0.70 | 20 |

The peak values and 3D distributions of antenna heat load on different plates are depicted in Figure 13 and Figure 14, respectively. The main results can be summarized as follows:

1. In general, with gas puffing, the peak values on all antenna plates increase in all cases that can be as much as 2 to 3 times. The gas puffing primarily provides a particle source in the far-SOL region, thereby increasing/flattening, e.g. in Figure 15(c), the density distribution within the SOL. With fixed thermal conductivity across the field, higher SOL densities lead to an increase in the radial heat flux, which in turn results in a higher heat load on the antenna surfaces.
2. The details of heat load performances depend on gas puffing locations. As the puffing positions move along the toroidal direction as in Case A and B, the peak values have no evident differences at side plate1. This feature is determined by the relative positions of 3D antenna structure and the gas puffing position and the SOL magnetic field geometry. This can be illustrated by the electron density distribution in Figure 15 and Figure 16. The injected deuterium natural particles are ionized when they enter SOL and the resulting ions are transported along the magnetic field lines in both the positive toroidal Φ+ and negative toroidal Φ− directions. When the ions move in the Φ− direction, most of them are directly far away from the side plate1 without plasma-wall interaction along magnetic field line and less of the ions deposited on side plate1 as in Figure 15 (a) and Figure 16 (a). The density that surrounds the side plate1 is comparable in Case A and B as shown in Figure 15 (a) (c) and Figure 16 (a)(c). Thus, the heat loads on side plate1 are similar. It is worth noticing that the plasma domain is not extended to the first wall as in Figure 4 (b), accordingly there is no plasma-wall interaction for Z<0 region of side plates as in Figure 4 (b). As far as the heat load on the side plate2 is concerned, the value of Case A is higher than Case B which is consistent with density distributions near the antenna in Figure 15 (c) and Figure 16 (c). As the gas puffing position moves along the toroidal direction from 20° to 40° as in Case A and C, the relative distance between the gas puffing position and side plate2 increases and fewer ions travel to the region near side plate2.

3. When the puffing positions are changed vertically as in Case A, C and D, the peak values at side plate1 respectively increase up to 3.8 to 6.9 and 8.5 $MWm^{-2}$ as summarized in Figure 13. This happens because when the gas puffing position moves upward, the injected particles that, after being ionized, flow along the magnetic field lines are less effectively shadowed by side plate2 and are thus transported toward the vicinity of side plate1 as Figure 15(a), Figure 17(a) and Figure 18(a). This explains why Case D exhibits a lower peak heat load on side plate2 compared to Case C: as the puffing location shifts further upward, the plasma transport resulting from gas injection hits the top and front plates instead of side plate2, as shown in Figure 17 (c) and Figure 18(c). This result also provide insight into a trade-off between ICRH power coupling and antenna heat load protection: on the one hand, gas puffing increases the localized density in the vicinity of the antenna, which is beneficial for power coupling [15]; on the other hand, the higher density could lead to a significantly increase of heat load on the antenna surfaces.
4. As to the top plate, all gas puffing cases lead to an increase in the heat load compared to the reference case; Regarding the front plate, gas puffing leads to a larger heated area, while the peak value of the heat load remains nearly unchanged, at about 0.2-0.3 $MWm^{-2}$ as can been appreciated in Figure 14.
5. The 3D gas puffing results in localized density spots in the far-SOL region, as illustrated in the poloidal cross-sectional views at $\phi$=0° in panels (b) of Figure 15-18. Two distinct types of density spots are identified: (1) the one in the upper region (approximately R~1.8m, Z~1.4m), highlighted by a red rectangular, which is attributed to ionized particles from gas puffing transported along the positive toroidal direction $\Phi$+; (2) the other in the region between the outer midplane (OMP) and the divertor entrance, identified by blue rectangle which is attributed to ionized particles from gas puffing transported along the positive toroidal direction $\Phi$-.

The above results demonstrate, for the first time, the successful capture of the coupled effects of three-dimensional gas puffing and geometry in DTT 3D edge plasma modelling. There are still uncertainties in the simulations: 1) gas puffing could lead to the changes of localized transport coefficients while the constant values are used in this study; 2) the DTT gas puffing system is still being designed, the current gas puffing strategy, e.g. the rate and position will be optimized; 3) The plasma domain does not extend to the first wall that the gas puffing point source and antenna structure are not fully covered by the plasma computational domain. In future work, these uncertainties on the antenna heat load will be investigated.

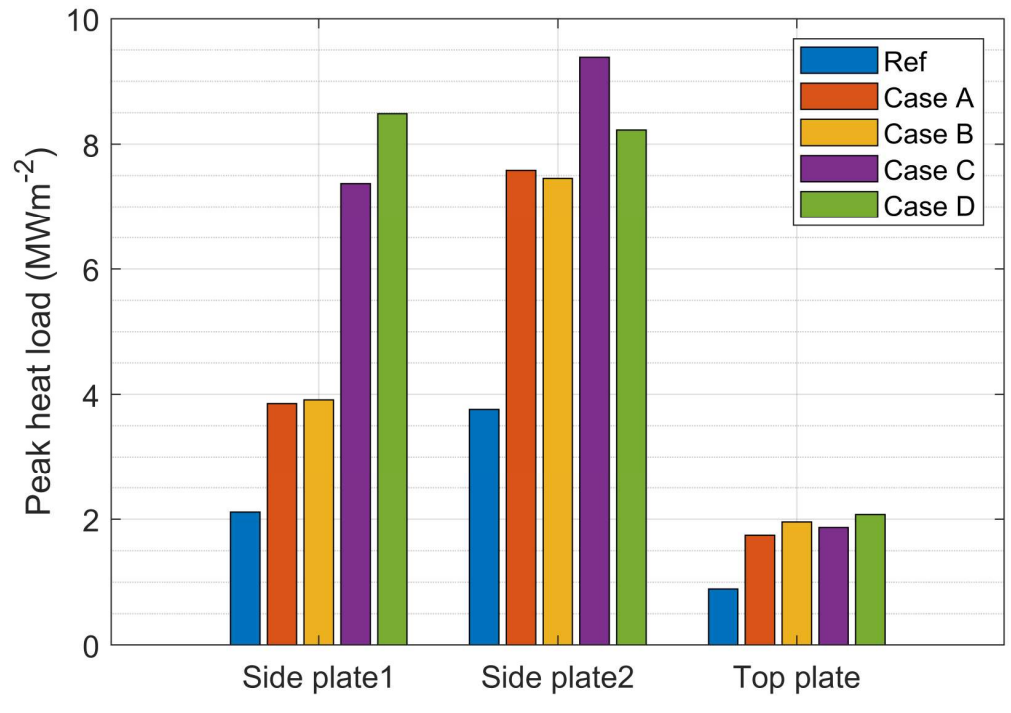

*Figure 13 Peak values of antenna heat load on Side plate1, Side plate2 and Top plate of the cases with different gas puffing positions. The gap puffing locations in the four cases are in Table 2.*

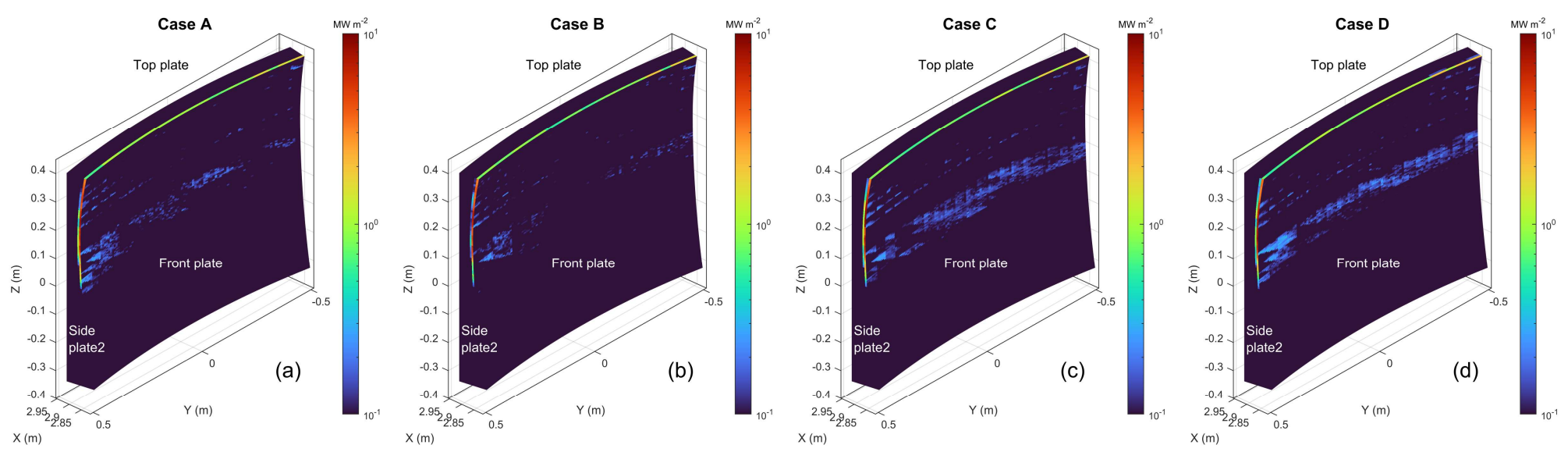

*Figure 14 ICRH Antenna 3D heat load distributions for: (a) Case A; (b) Case B; (c) Case C and (d) Case D. The gap puffing locations in the four cases are in Table 2. A log scale is used for the color bar to make the front plate pattern clearer.*

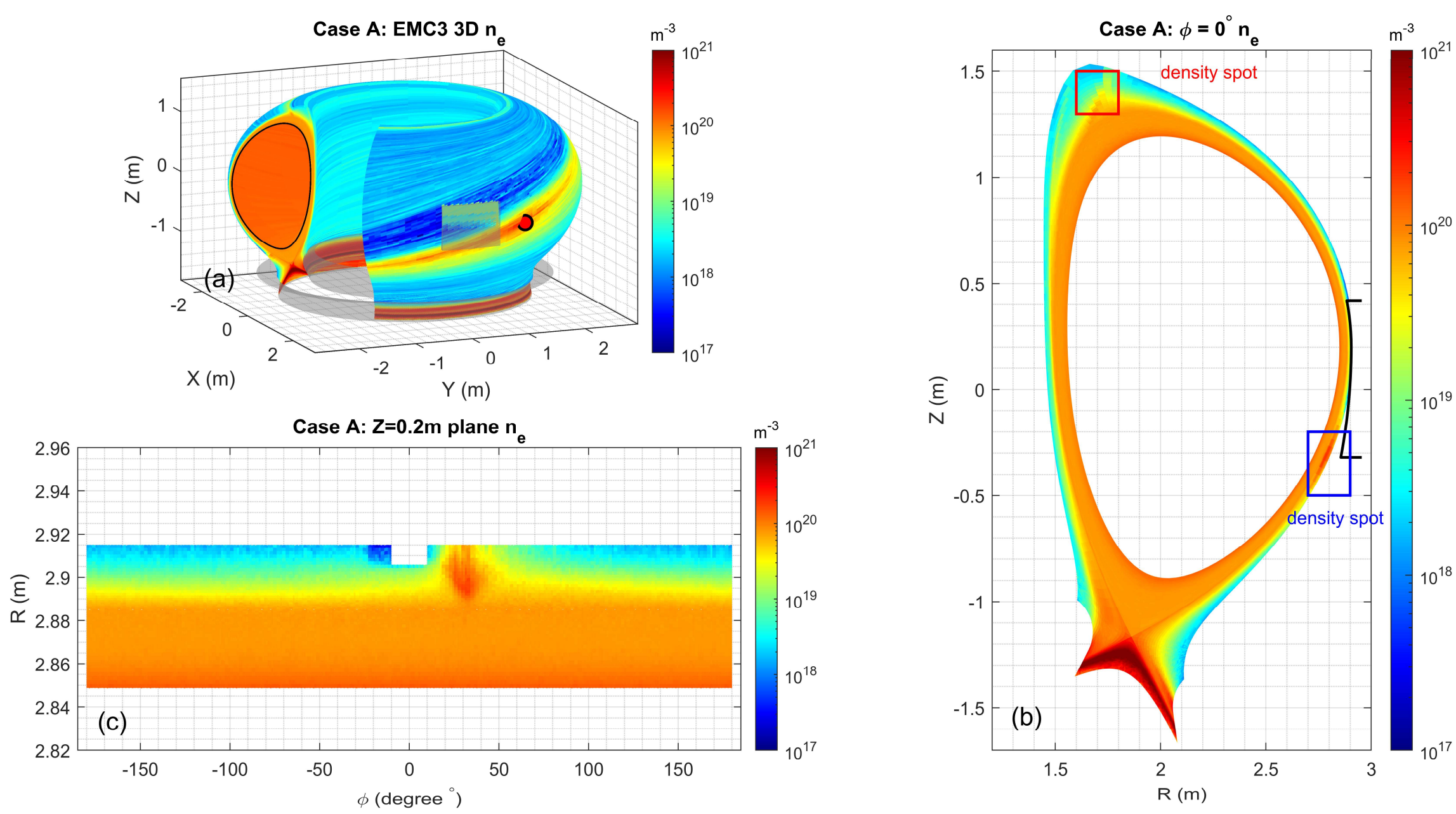

*Figure 15 Electron density distributions of Case A: (a) 3D view with gas puffing position which is R=2.95m Z=0.00m and Φ=20° marked by a red spot with a black circle; (b) poloidal cross-section view at ϕ=0° and (c) toroidal view at Z=0.2 m.*

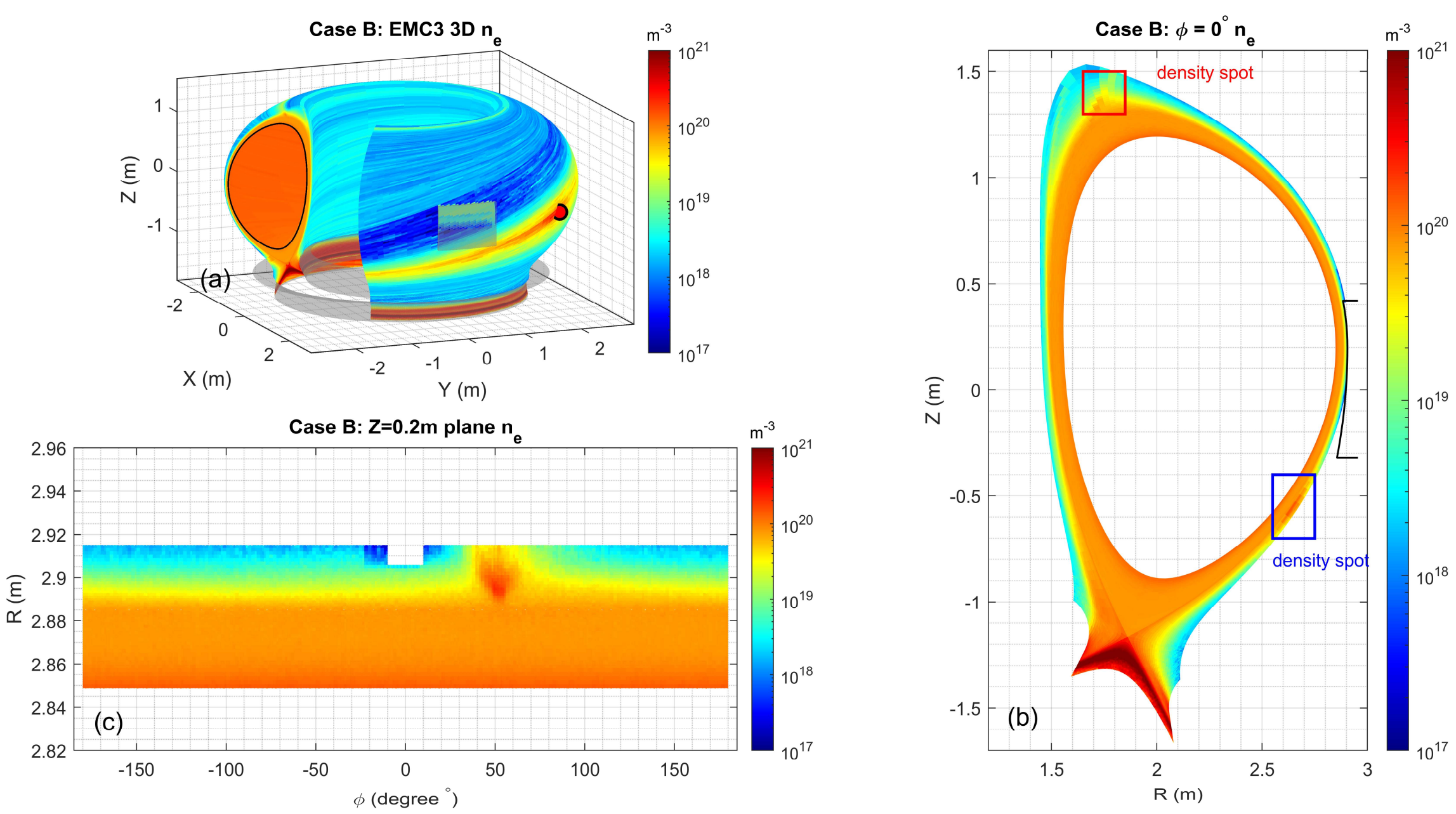

*Figure 16 Electron density distributions of Case B: (a) 3D view with gas puffing position which is R=2.95m Z=0.00m and Φ=40° marked by a red spot with a black circle; (b) poloidal cross-section view at ϕ=0° and (c) toroidal view at Z=0.2 m.*

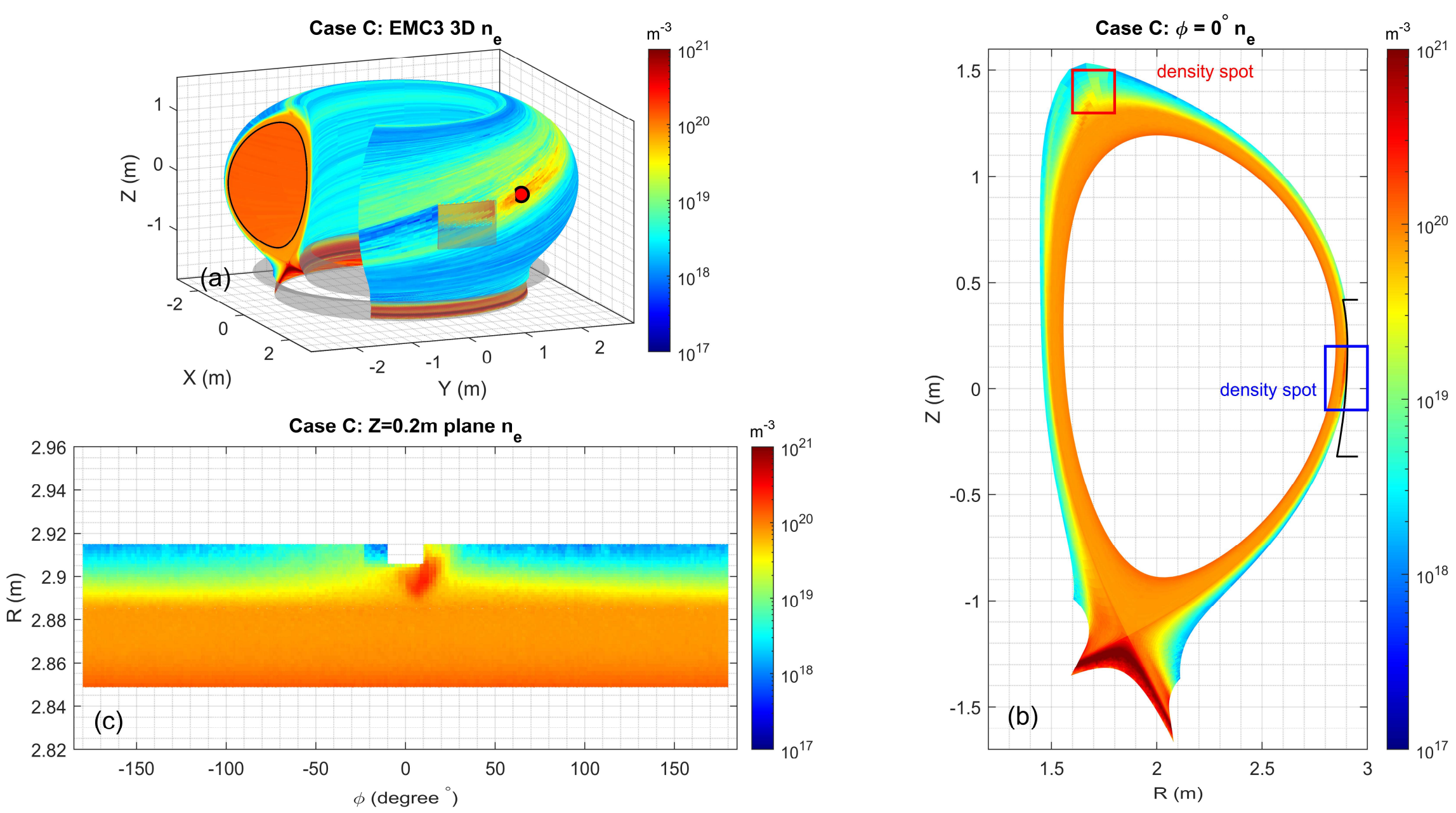

*Figure 17 Electron density distributions of Case C: (a) 3D view with gas puffing position which is R=2.95m Z=0.45m and Φ=20° marked by a red spot with a black circle; (b) poloidal cross-section view at ϕ=0° and (c) toroidal view at Z=0.2 m.*

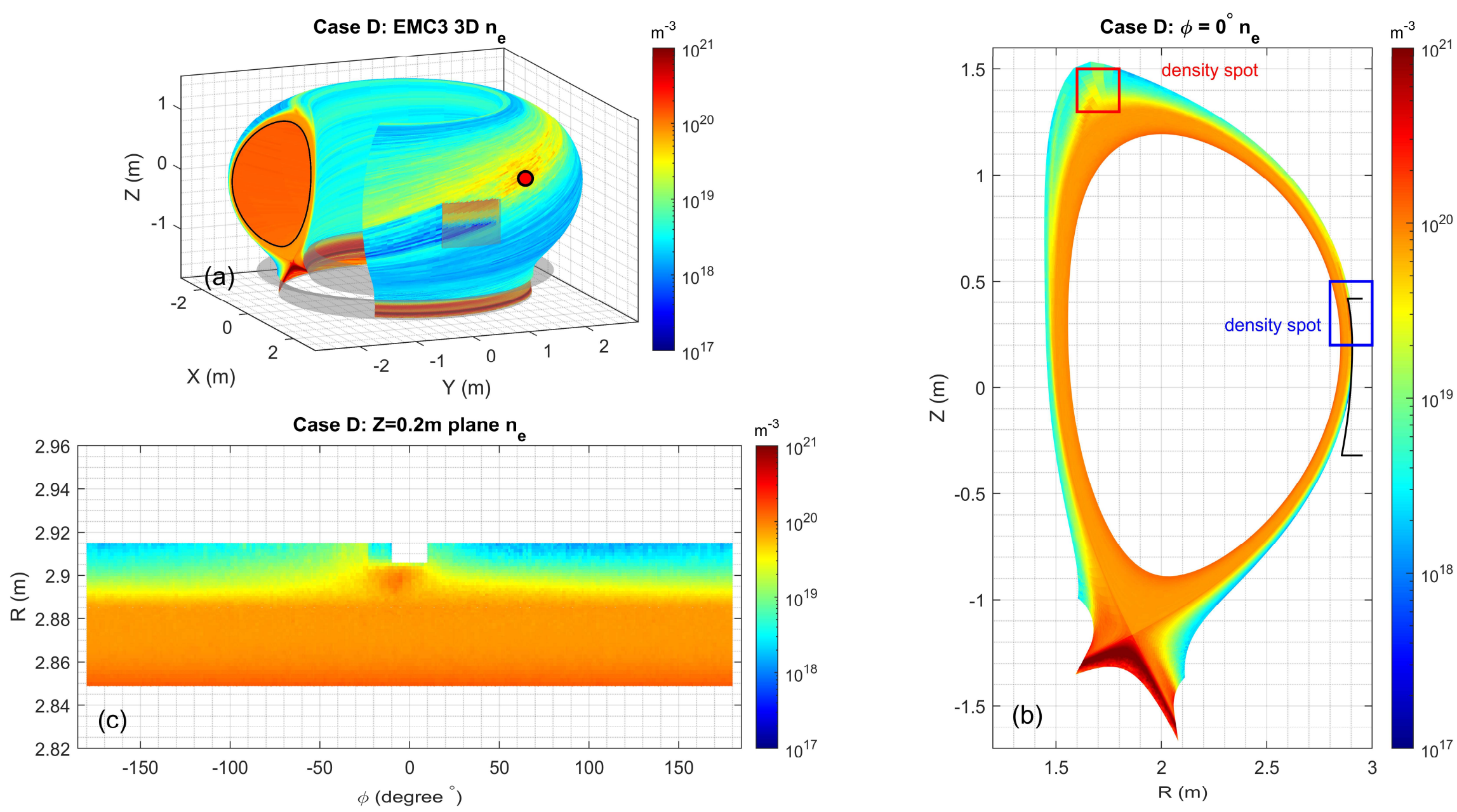

*Figure 18 Electron density distributions of Case D: (a) 3D view with gas puffing position which is R=2.95m Z=0.45m and Φ=20° marked by a red spot with a black circle; (b) poloidal cross-section view at ϕ=0° and (c) toroidal view at Z=0.2 m.*

## 7. Summary and outlook

The three-dimensional (3D) edge plasma transport code EMC3-EIRENE, for the first time, has been applied on the Divertor Tokamak Test (DTT) facility, which is aimed at assessing the thermal load on the ICRH antenna that supports the design of the ICRH antenna system. As a first step towards ensuring the reliability of the predicted results, we performed a benchmark of EMC3-EIRENE 3D simulations against SOLPS-ITER 2D simulations under toroidally symmetric conditions, in the absence of the 3D antenna structure. Good agreements of OMP and OT profiles are obtained especially in the far-SOL region, where the antenna structure is located. This indicates that the EMC3-EIRENE modeling setup is reliable and suitable for investigating 3D effects.

Subsequently, the 3D antenna structure was incorporated into the DTT high-power scenario modelling. A parametric scan of antenna toroidal symmetry, which represents the number of antennas in the full toroidal domain, was performed. It was found that as the antenna toroidal symmetry decreases, the peak heat loads on the antenna plates increase due to the deposition of power over less total area. A simulation strategy was proposed in which toroidal symmetry=3 is preferable compared to toroidal symmetry=1, as it substantially reduces the computational cost by considering only one-third of the toroidal domain, while incurring only a minor loss of accuracy. A reliable 3D antenna heat load distribution was obtained, with peak values on the front plate, top plate, side plate1, and side plate2 of 0.2, 0.9, 2.1, and 3.8 MW/m$^2$, respectively. A scan of perpendicular transport coefficients reveals that the heat load was influenced by the plasma density. Increasing the transport coefficients leads to a corresponding increase in the heat load up to 1-2 order of magnitude.

Finally, 3D gas puffing was included in the modelling to investigate its effect on the antenna heat load. The coupled effects of three-dimensional gas puffing and the structural geometry are successfully captured in DTT edge plasma modelling. Gas puffing primarily provides a particle source in the far-SOL region, thereby increasing/flattening the density distribution within the SOL. If the thermal conductivity across the field is fixed, as is likely the cases in this study, higher SOL densities lead to an increase in the radial heat flux, which in turn results in a higher thermal load on the antenna, although the details depend on the gas puff location. The corresponding 3D plasma behavior was presented, showing that the plasma density at the antenna plates is determined by the relative positions between the gas puffing and the antenna structure, together with the magnetic geometry.

As the first application of EMC3-EIRENE on DTT, this work provides a solid foundation for future studies. Several aspects of this work remain to be improved: (1) The current plasma computational domain is not extended to the first wall; therefore the antenna structure is not fully covered by the plasma. Full-domain modeling is foreseen in the future; (2) Impurity seeding is not considered in the present study and will be incorporated in future work; (3) Volumetric recombination is not considered in the present study, as the focus is on upstream performance rather than divertor detachment. This will be included in future work to investigate divertor performance under 3D effects; (4) Constant transport coefficients are adopted in the current study and will be replaced by physically based non-uniform transport coefficients [31] in future investigations; (5) This study focuses on the standard single-null divertor configuration. In future work, 3D edge plasma modeling will be extended to DTT advanced divertor configurations [28], involving negative triangular divertor, X-divertor, and snowflake divertor configurations.

## Acknowledgement

This work is supported by PNRR M4C2—HPC, Big data and Quantum Computing (Simulazioni, calcolo e analisi dei dati e altre prestazioni—CN1)—CUP I53C22000690001 SPOKE 6 Multiscale modelling & Engineering applications.

## References

[1]. F. Romanelli et al., “Divertor Tokamak Test facility project: status of design and implementation,” Nuclear Fusion, vol. 64, no. 11, p. 112015, Sep. 2024, doi: 10.1088/1741-4326/ad5740.

[2]. F. Crisanti et al., “Physics basis for the divertor tokamak test facility,” Nuclear Fusion, vol. 64, no. 10, p. 106040, Sep. 2024, doi: 10.1088/1741-4326/ad6e06.

[3]. S. Ceccuzzi et al., “Progress in the development of the ICRF system of DTT,” Fusion Engineering and Design, vol. 213, p. 114849, Apr. 2025, doi: 10.1016/j.fusengdes.2025.114849.

[4]. G. Camera, F. G. Lanzotti, S. Ceccuzzi, and G. D. Gironimo, “Preliminary Conceptual Design of the ICRH Antenna for DTT: A Systems Engineering Approach,” IEEE Transactions on Plasma Science, vol. 52, no. 9, pp. 4075–4081, Sep. 2024, doi: 10.1109/TPS.2024.3369469.

[5]. M. Firdaouss, V. Riccardo, V. Martin, G. Arnoux, and C. Reux, “Modelling of power deposition on the JET ITER like wall using the code PFCFLux,” Journal of Nuclear Materials, vol. 438, pp. S536–S539, Jul. 2013, doi: 10.1016/j.jnucmat.2013.01.111.

[6]. L. Kos, R. A. Pitts, G. Simič, M. Brank, H. Anand, and W. Arter, “SMITER: A field-line tracing environment for ITER,” Fusion Engineering and Design, vol. 146, pp. 1796–1800, Sep. 2019, doi: 10.1016/j.fusengdes.2019.03.037.

[7]. Y. Feng, F. Sardei, J. Kisslinger, and P. Grigull, “A 3D Monte Carlo code for plasma transport in island divertors,” Journal of Nuclear Materials, vol. 241–243, pp. 930–934, Feb. 1997, doi: 10.1016/S0022-3115(97)80168-7.

[8]. Y. Feng et al., “Recent Improvements in the EMC3-Eirene Code,” Contributions to Plasma Physics, vol. 54, no. 4–6, pp. 426–431, Jun. 2014, doi: 10.1002/ctpp.201410092.

[9]. D. Reiter, M. Baelmans, and P. Börner, “The EIRENE and B2-EIRENE Codes,” Fusion Science and Technology, vol. 47, no. 2, pp. 172–186, Feb. 2005, doi: 10.13182/FST47-172.

[10]. Y. Feng, F. Sardei, J. Kisslinger, P. Grigull, K. McCormick, and D. Reiter, “3D Edge Modeling and Island Divertor Physics,” Contributions to Plasma Physics, vol. 44, no. 1–3, pp. 57–69, Apr. 2004, doi: 10.1002/ctpp.200410009.

[11]. Y. Feng et al., “Understanding detachment of the W7-X island divertor,” Nuclear Fusion, vol. 61, no. 8, p. 086012, Jun. 2021, doi: 10.1088/1741-4326/ac0772.

[12]. H. Frerichs, D. Reiter, O. Schmitz, T. E. Evans, and Y. Feng, “Three-dimensional edge transport simulations for DIII-D plasmas with resonant magnetic perturbations,” Nuclear Fusion, vol. 50, no. 3, p. 034004, Feb. 2010, doi: 10.1088/0029-5515/50/3/034004.

[13]. T. Lunt et al., “EMC3-Eirene simulations of particle- and energy fluxes to main chamber- and divertor plasma facing components in ASDEX Upgrade compared to experiments,” Journal of Nuclear Materials, vol. 463, pp. 744–747, 2015, doi: https://doi.org/10.1016/j.jnucmat.2014.09.020.

[14]. L. Scotti et al., “Assessing wall heat fluxes in the quasi-continuous exhaust regime at ASDEX Upgrade,” Nuclear Fusion, vol. 66, no. 4, p. 046008, Mar. 2026, doi: 10.1088/1741-4326/ae4bf5.

[15]. W. Zhang et al., “3D simulations of gas puff effects on edge density and ICRF coupling in ASDEX Upgrade,” Nuclear Fusion, vol. 56, no. 3, p. 036007, Feb. 2016, doi: 10.1088/0029-5515/56/3/036007.

[16]. M. Brank et al., “Assessment of plasma power deposition on the ITER ICRH antennas,” Nuclear Materials and Energy, vol. 27, p. 101021, Jun. 2021, doi: 10.1016/j.nme.2021.101021.

[17]. H. Frerichs, “FLARE: field line analysis and reconstruction for 3D boundary plasma modeling,” Nuclear Fusion, vol. 64, no. 10, p. 106034, Sep. 2024, doi: 10.1088/1741-4326/ad7303.

[18]. X. Bonnin, et al., “Presentation of the New SOLPS-ITER Code Package for Tokamak Plasma Edge Modelling,” Plasma and Fusion Research, vol. 11, pp. 1403102–1403102, 2016, doi: 10.1585/pfr.11.1403102.

[19]. S. Wiesen et al., “The new SOLPS-ITER code package,” Journal of Nuclear Materials, vol. 463, pp. 480–484, Aug. 2015, doi: 10.1016/j.jnucmat.2014.10.012.

[20]. Feng, Y., et al. "EMC3-Eirene/SOLPS4. 3 comparison for ITER." *38th EPS Conference on Plasma Physics*. European Physical Society, 2011.

[21]. R. Marchand and M. Dumberry, “CARRE: a quasi-orthogonal mesh generator for 2D edge plasma modelling,” Computer Physics Communications, vol. 96, no. 2, pp. 232–246, Aug. 1996, doi: 10.1016/0010-4655(96)00052-5.

[22]. G. Giruzzi et al., “The Divertor Tokamak Test facility research plan,” Nuclear Fusion, vol. 66, no. 11, p. 116017, Jul. 2026, doi: 10.1088/1741-4326/ae7a8d.

[23]. V. Kotov et al., “Numerical modelling of high density JET divertor plasma with the SOLPS4.2 (B2-EIRENE) code,” Plasma Physics and Controlled Fusion, vol. 50, no. 10, p. 105012, Sep. 2008, doi: 10.1088/0741-3335/50/10/105012.

[24]. H. Frerichs et al., “Stabilization of EMC3-EIRENE for detachment conditions and comparison to SOLPS-ITER,” Nuclear Materials and Energy, vol. 18, pp. 62–66, 2019, doi: https://doi.org/10.1016/j.nme.2018.11.022.

[25]. Y. Feng, D. Reiter, H. Frerichs, and the W7-X Team, “A novel method for treating MAR in EMC3-Eirene, and first applications to W7-X,” Nuclear Fusion, vol. 65, no. 6, p. 066008, May 2025, doi: 10.1088/1741-4326/add27a.

[26]. D. Harting, S. Wiesen, H. Frerichs, D. Reiter, P. Börner, and Y. Feng, “Validating the 3D edge code EMC3-EIRENE against 2D simulations with EDGE2D-EIRENE for JET single null configurations,” Journal of Nuclear Materials, vol. 415, no. 1, Supplement, pp. S540–S544, Aug. 2011, doi: 10.1016/j.jnucmat.2011.01.030.

[27]. S. Ceccuzzi et al., “The ICRF antenna of DTT: Design status and perspectives,” AIP Conference Proceedings, vol. 2984, no. 1, p. 030015, Aug. 2023, doi: 10.1063/5.0162417.

[28]. P. Innocente et al., “Design of a multi-configurations divertor for the DTT facility,” Nuclear Materials and Energy, vol. 33, p. 101276, 2022, doi: https://doi.org/10.1016/j.nme.2022.101276.

[29]. W. Dekeyser et al., “Plasma edge simulations including realistic wall geometry with SOLPS-ITER,” Nuclear Materials and Energy, vol. 27, p. 100999, Jun. 2021, doi: 10.1016/j.nme.2021.100999.

[30]. W. Zhang et al., “Parametric study of midplane gas puffing to maximize ICRF power coupling in ITER,” Nuclear Fusion, vol. 63, no. 3, p. 036008, Feb. 2023, doi: 10.1088/1741-4326/acb4ad.

[31]. L. Balbinot et al., “Multi-code estimation of DTT edge transport parameters,” Nuclear Materials and Energy, vol. 34, p. 101350, 2023, doi: https://doi.org/10.1016/j.nme.2022.101350.